\documentclass[journal,comsoc]{IEEEtran}
\usepackage{config}
\newcommand{\subscript}[2]{$#1 _ #2$}

\begin{document}
\title{Declarative Traffic Engineering for\\ Low-Latency and Reliable Networking}

 \author{Jacopo Massa,
 Stefano Forti,
 Federica Paganelli,
 Patrizio Dazzi,
 Antonio Brogi,
 Alexander Clemm,
 Toerless Eckert
 \thanks{J. Massa, S. Forti, F. Paganelli, P. Dazzi and A. Brogi are with the Department of Computer Science, University of Pisa, 
 Pisa, Italy}
 \thanks{A. Clemm is with Sympotech, Los Gatos, California/USA.}
 \thanks{T. Eckert is with Futurewei, USA.}
 }


\maketitle

\begin{abstract}

Cloud-Edge applications like industrial control systems and connected vehicles demand stringent end-to-end latency guarantees. Among existing data plane candidate solutions for bounded latency networking, the guaranteed Latency-Based Forwarding (gLBF) approach ensures punctual delivery of traffic flows by managing per-hop delays to meet specific latency targets, while not requiring that per-flow states are maintained at each hop. 
However, as a forwarding plane mechanism, gLBF does not define the control mechanisms for determining feasible forwarding paths and per-hop latency budgets for packets to fulfil end-to-end latency objectives.  In this work, we propose such a control mechanism implemented in Prolog that complies with gLBF specifications, called declarative gLBF (\toolname). 
The declarative nature
 of Prolog allows our prototype to be concise ($\simeq 120$ lines of code) and easy to extend. We show how the core dgLBF implementation is extended to add reliability mechanisms, path protection, and fate-sharing avoidance to enhance fault tolerance and robustness.
 Finally, we evaluate the system's performance through simulative experiments under different network topologies and with increasing traffic load to simulate saturated network conditions, scaling up to 6000 flows. Our results show a quasi-linear degradation in placement times and system resilience under heavy traffic. 

\end{abstract}


\section{Introduction}
\label{sec:introduction}

The increasing adoption of the Internet of Things (IoT) applications in mission-critical domains, such as connected vehicles, Industry 4.0, and remote surgery, has brought new challenges to networking systems. These applications demand stringent end-to-end latency guarantees, requiring communication delays to either remain below a strict threshold (\textit{in-time} service) or occur within a specific time window (\textit{on-time} service)~\cite{itu}. However, traditional network systems struggle to meet these deterministic latency requirements while maintaining performance and scalability.

Emerging paradigms like Cloud-Edge computing~\cite{ce-survey} and the transition toward 6G networks~\cite{6gsurvey} underscore the growing need for precise latency control. Mission-critical systems demand strict adherence to latency targets, unlike conventional approaches where latency is reduced to \textit{low} or \textit{ultra-low} levels. A missed target can lead to severe failures, as seen in teleoperated robotic surgeries or industrial automation systems, where ironclad guarantees are required instead~\cite{ClemmTNSM23}.

Since traditional ``best-effort'' solutions, which rely on statistical properties, cannot guarantee the service levels required by latency-sensitive networking applications, high-precision networking has gained significant attention both in the research community and in standardization forums (e.g, Time-Sensitive Networking (TSN) \cite{tsn} at Layer 2, and more recently, Deterministic Networking (DetNet) \cite{detnet}  at Layer 3). Among the candidate solutions for DetNet networking, the \textit{guaranteed Latency-Based Forwarding} (gLBF) forwarding mechanism and latency calculus framework addresses these challenges by associating each packet with a latency budget, which guides forwarding decisions at each hop~\cite{lbf, glbf} (e.g. delaying, forwarding the packet immediately or dropping it when the latency target cannot be met). Indeed, this budget reflects the total latency target, the latency already experienced, and the expected delay along the remaining path. By dynamically managing these budgets, gLBF ensures latency objectives are met without the overhead of maintaining a per-flow state at every node. Yet, a critical challenge remains: selecting paths and configuring delays to meet these latency goals efficiently.

To cope with this challenge, we propose adopting a declarative approach, leveraging Prolog, to implement control operations for gLBF deployments. This approach would determine paths for packet forwarding and compute latency budgets for packets, accounting for latency objectives and path characteristics (e.g., hops, propagation delay, and bandwidth capacity).

Prolog is a logic programming language where programs consist of \textit{facts} and \textit{rules}, which can be used to define solutions to a given problem rather than specifying the procedural steps to solve it. 
Additionally, Prolog supports a concise notation that can be easily extended to accommodate new requirements by adding more facts and rules \cite{warren2023prolog}.

Indeed, high-precision networking applications often come with additional requirements. First, an application may necessitate selecting a path that guarantees a minimum level of path reliability, where path reliability refers to the probability that the entire path remains functional. This requires monitoring and diagnostic capabilities to assess the reliability of the individual links that make up the path.
Second, a higher level of fault tolerance may be needed, which can be achieved by implementing redundancy forwarding mechanisms, such as 1+1 protection. This mechanism replicates traffic flows over at least two disjoint paths, ensuring service continuity in case of failure.
Third, in some application domains, ensuring that two flows, such as application data flows and control flows, do not share the same path is essential. For example, symmetric key-encrypted data flows are quantum safe, but negotiating symmetric keys across the control plane may not be able to use quantum safe crypto. Making it more difficult to observe data and control flows partially mitigates this challenge.

Thus, the main contribution of this work consists in:
\begin{enumerate}
\item The proposal of \textit{declarative gLBF} (\toolname), a concise and flexible Prolog-based implementation of control operations for gLBF (i.e., flows placement and delay configuration). The core implementation of \toolname has been introduced in a previous work ~\cite{wscc}. By leveraging Prolog’s declarative programming, \toolname provides an extremely simple yet easy to customise control-plane, focusing on what needs to be achieved rather than prescribing how to achieve it. Simplicity makes \toolname particularly suitable for dynamic Cloud-Edge environments to reduce opex. Customisation becomes important when network topologies become more complex and application requirements grow more varied and stringent in reliability and fault tolerance.
\item Prior work \cite{wscc} is further extended by introducing key enhancements to address the above-mentioned challenges:
\begin{enumerate*}[label=\textit{(\roman{*})}]
    \item reliability constraints to evaluate path reliability based on the probability of link failures, ensuring resilient and dependable communication;
    \item a 1+1 path protection strategy that assigns disjoint paths to each flow, safeguarding against service interruptions caused by link failures;
    \item anti-affinity policies to avoid fate-sharing, ensuring that distinct flows -- such as data and control -- are routed independently to minimise simultaneous failure risks.
\end{enumerate*}
\end{enumerate}

We believe that with these enhancements, \toolname is better equipped to handle the demands of large-scale, real-world networks. 
Our prototype accepts detailed network infrastructure and flow requirements descriptions, computes 
paths and determines the appropriate per-hop delays to guarantee stringent latency objectives. We tested \toolname on realistic and synthetic network topologies, demonstrating its scalability and effectiveness. The results confirm its suitability for mission-critical applications, where deterministic latency guarantees must be maintained without compromising reliability or performance.

The rest of this paper is structured as follows: \cref{sec:background} reviews gLBF and its foundational principles. \Cref{sec:methodology} presents the declarative framework and the added reliability features. \Cref{sec:example} discusses the experimental evaluations, while~\cref{sec:scalability} highlights scalability insights. \Cref{sec:related} explores related literature, and~\cref{sec:conclusions} concludes with potential directions for future work.
\section{Background}
\label{sec:background}

This section provides background information on past solutions for internetworking latency guarantees.

\paragraph{IETF QoS}
There is a long history of past solutions to try and overcome the limitations of the IP best-effort service.
Of particular note are the IETF's efforts to provide QoS in IP networks, which have produced different architectures: Differentiated Services (DiffServ) and Integrated Services (IntServ).
DiffServ \cite{rfc2475} addresses the need for simple and coarse-grained QoS by
dividing packets into classes by marking the DSCP field in the IP header. This allows specific per-hop forwarding behaviours to be applied to packets along their path. 
Intserv, originally defined in the IETF RFC 1633 \cite{rfc1633}, encompasses two primary classes of service: Guaranteed Service and Controlled-Load service. In particular, the Guaranteed Service (GS) class provides a stringent QoS guarantee, ensuring per-flow bandwidth guarantees, strict bounds on end-to-end delay, and no packet loss \cite{rfc2212}. IntServ uses admission control and resource reservation mechanisms that involve the intermediate routers in the definition of end-to-end paths. 
A primary drawback is using per-flow state and per-flow processing at the routers, resulting in greater complexity and hardware expense while leading to scalability problems in large networks. Moreover, GS does not offer mechanisms to slow down packets based on their desired minimum latency \cite{ClemmTNSM23}.

\paragraph{Time-Sensitive Networking}
Below the network layer (L3), Time-Sensitive Networking (TSN) \cite{tsn} is a set of standards developed by the IEEE 802.1 working group to enable time-sensitive data transmission over Ethernet networks. TSN aims to fulfil 
 professional applications' requirements over local area networks (LANs) with priority queuing, preemption, and traffic shaping mechanisms. 
 Among the TSN standards, ATS (Asynchronous Traffic Shaper) specifies mechanisms like IntServ GS do not require clock synchronisation between network nodes but continue to require per-hop, per-flow state. ATS state and packet processing is simpler than GS but at the expense of less stringent latency guarantees for the same scenarios. ATS primarily optimises smaller-scale topologies and fewer traffic flows, conditions typically found in industrial networking local area network environments.

\paragraph{IETF DetNet}
To integrate TSN technologies into IP networks, the IETF DetNet (Deterministic Networking) working group has defined a general Deterministic Networking architecture \cite{rfc8655}. The DetNet architecture is designed to provide bounded latency and low packet loss for time-sensitive applications over IP networks under a single administrative control, such as private WANs and campus networks. However, DetNet does not specify or mandate a particular solution within that framework. One candidate solution is guaranteed Latency-Based Forwarding (gLBF), which possesses several important characteristics, such as the fact that it does not require nodes to maintain per-flow state. This reduces hardware cost (as less memory is required), solution complexity, and deployment cost. While gLBF could equally be used in an L2 TSN, it was primarily targeted for DetNet contexts after evaluations, where TSN solutions, including ATS, were seen as too complex for large-scale DetNet deployments. 

\paragraph{Latency-Based Forwarding and gLBF}
gLBF \cite{glbf} evolved from an earlier technology, Latency Based Forwarding (LBF) \cite{lbf}. LBF's basic idea is as follows: a latency objective for a flow is given, which can be carried as a piece of metadata as part of the packet. Combined with knowledge of the packet's path (including the number of hops and link propagation delays), a latency budget can be calculated that determines how much time the packet can afford to spend across the path for processing. From this, each hop computes a budget for the packet's dwell time and applies appropriate queuing decisions using push-in-first-out (PIFO) queues that ensure the dwell time is met.  

%
LBF ensures that packets meet their latency objective when no packet collisions occur, including delaying packets that are forwarded faster than necessary when they contend with more urgent packets on a link. However, LBF forwarding was defined so flexibly that no mathematical model could easily calculate whether a particular set of flows could be admitted to the network under high loads. Hence, LBF is today best suited only for non-deterministic deployments.
gLBF resolves this limitation by modifying and simplifying the LBF forwarding rules so that the well-known
ATS latency calculus can be re-used for deterministic deployment while still maintaining the stateless forwarding
nature of LBF.
When a node receives a packet, the packet's further processing is delayed until the previous node's packet dwell time budget is entirely exhausted. While it may seem counter-intuitive initially to slow down packets, this way, burst aggregation is avoided by ensuring that packets remain precisely spaced while still ensuring that packets remain within their acceptable latency bounds. Accordingly, guarantees for packets with specific latency objectives can finally be given. Importantly, required state information (latency objective and previous node's dwell time) is carried within the packet, obviating the need to maintain a separate flow state at the node. 

However, for solutions to be deployed, there is still one critical component that is missing: the paths for packet forwarding need to be determined, and latency budgets for packets still need to be computed based on latency objectives and the characteristics of the path (e.g., the hops and their propagation delays). This is the missing piece which is addressed in this paper. The system presented in this paper allows network operators to specify a set of desired outcomes (i.e., a set of flows that need to be accommodated with their respective latency objectives). The system subsequently identifies the paths to which the flows are assigned and determines the latency budgets to which gLBF is applied.  We call this solution dgLBF (declarative gLBF).
\section{Declarative methodology}
\label{sec:methodology}

\noindent
In this section, we first give some needed background on the Prolog programming language (\Cref{sec:prolog}). Then, we detail the knowledge base used by our prototype  (\Cref{sec:kr}) and how it implements a declarative executable specification of gLBF~\cite{glbf} control operations considering latency guarantees (\Cref{sec:vanilla_dglbf}). Subsequently, we illustrate extensions for: 
\begin{enumerate*}[label=\textit{(\roman{*})}]
    \item guaranteeing flow placements for minimum path reliability by introducing per link reliabilities (\cref{sec:linkrel}),
    \item avoiding fate sharing between flows (\cref{sec:nofatesharing}), and
    \item enabling path protection (\cref{sec:pathprotection}).
\end{enumerate*}

\subsection{Background: Prolog}
\label{sec:prolog}
Prolog programs consist of \textit{clauses} of the form \cd{a :- b1, ..., bn.} stating that \cd{a} holds if \cd{b1} $\wedge \ldots \wedge$ \cd{bn} holds. Clauses with empty premise (\cd{n}$=0$) are called \textit{facts}. Predicate definitions can also contain \textit{disjunctions} (denoted by \cd{;}) and negations (denoted by \cd{\textbackslash+}). Variables start with upper-case letters, and square brackets denote lists (e.g. \cd{[L|Ls]}, with \cd{L} representing the first element and \cd{Ls} the rest of the list).
A Prolog \textit{predicate} is represented in the form \cd{predicate/arity}, where \textit{predicate} is the predicate's name and \textit{arity} is the number of arguments it takes. For example, \cd{city/2} indicates a predicate \cd{city} with two arguments.
Prolog programs can be queried, and the Prolog interpreter tries to answer each query by applying \textit{Selective Linear Definite} (SLD) resolution~\cite{lloyd} and by returning a computed answer substitution instantiating the variables in the query. In this article, we rely on the SWI-Prolog implementation~\cite{swi-prolog}. For instance, the query \cd{?- romantic(City).} on the program

\begin{code}
city(paris). 
city(pisa). 
city(london).

rainy(pisa). closeToTheSea(pisa).

romantic(City):- 
    city(City), 
    (rainy(City), closeToTheSea(City)) ; (City = paris).
\end{code}

\noindent
returns the computed answer substitutions

\begin{code}
?- romantic(City).
City = pisa ; City = paris.
\end{code}

\noindent rewriting the query by applying the first clause to retrieve a city non-deterministically, the second and third to check whether the retrieved city is rainy and close to the sea or it is Paris. This also shows the \textit{closed-world} assumption of Prolog. Indeed, London cannot be proved romantic according to the provided knowledge base and definition of \cd{romantic/1}.

\subsection{Knowledge representation}
\label{sec:kr}

The knowledge base of \toolname includes facts about infrastructure nodes, links and incoming flow requests.
Infrastructure \textit{nodes} are denoted by a unique identifier \cd{N} and characterised by their processing delay \cd{DProc}, i.e., the average latency incurred at the node for packet processing tasks, as in 

\begin{code}
node(N, DProc).
\end{code}

End-to-end \textit{links} between nodes \cd{N} and \cd{M} are described by their propagation delay \cd{DProp} (in ms), available bandwidth \cd{BW} (in Mbps) and reliability value \cd{LinkRel}, denoting the probability that such link is functional, through facts like

\begin{code}
link(N, M, DProp, BW, LinkRel).
\end{code}

\textit{Flow requests} starting from node \cd{SrcN} and ending in node \cd{DstN} are each denoted by a unique identifier \cd{F} as in

\begin{code}
flow(F, SrcN, DstN).
\end{code}

\noindent Each flow request \cd{F} also comes with a maximum acceptable end-to-end \cd{LatencyBudget} and the toleration threshold \cd{Th} for latency deviation (both expressed in ms). This implies that the latency between source and destination must stay within the range $[\cd{LatencyBudget} -\cd{Th}, \cd{LatencyBudget} +\cd{Th}]$. Last, each flow request includes key characteristics, such as packet size \cd{PktSize} (in Mb), burst size \cd{BstSize} (in number of packets), and bit \cd{Rate} (in Mbps):

\begin{code}
flowReqs(F, PktSize, BstSize, Rate, LatencyBudget, Th).
\end{code}

Reliability requirements for each flow request \cd{F} are defined in terms of the \cd{ReplicaFactor} of disjoint paths\footnote{\texttt{dgLBF} can generically manage $N$ replicas, even if we defined path protection as $1+1$ protected flow.} the required reliability \cd{ReqRel} $\in [0,1]$ for all replicated paths, as in:

\begin{code}
reliabilityReqs(F, ReqRel, ReplicaFactor).
\end{code}

Finally, we can specify that a given flow request \cd{F} should avoid sharing paths with other flows (i.e., \cd{AvoidedFlows}) through facts like

\begin{code}
antiAffinity(F, [AvoidedFlows]).
\end{code}

The non-deterministic behaviour of Prolog discourages its use for determining paths in large-scale networks. We, therefore, assume that \toolname inputs a set of candidate paths for each pair of source/destination nodes in the flows to be placed, which we precompute once by relying on the NetworkX Python library\footnote{Available at \url{https://networkx.org}. We rely on NetworkX v3.1, exploiting the function \fncd{all\_simple\_paths(G, src, dst)}, which in turn depends on a depth-first search to determine all eligible paths from node \fncd{src} to node \fncd{dst} in a graph \fncd{G}.}. 
Candidate end-to-end paths are made of infrastructure links and are denoted by a unique identifier \cd{P} and represented as the list of \cd{TraversedNodes}, starting from \cd{N} and ending in \cd{M}, as in
 
\begin{code}
candidate(P, N, M, TraversedNodes).
\end{code}

\subsection{Plain \toolname}
\label{sec:vanilla_dglbf}

\noindent The objective of the plain version of our Prolog reasoner is to determine which candidate paths can satisfy the latency requirements of flows to be placed and to compute all necessary network delays to be introduced along such a path. 
More detailedly, Given a set of flow requests to handle, the plain version of our \toolname prototype~\cite{wscc} performs the following two main consecutive steps:

\begin{enumerate}[label=(\subscript{S}{{\arabic*}})]
    \item\label{step1} the first step determines a suitable path for each incoming flow request by summing up processing and transmission delays along the path, and checking them against the \textit{minimum} latency budget (i.e., $\cd{LatencyBudget} - \cd{Th}$); it also computes needed delays through gLBF whenever flows could reach their destination ahead of time. Finally, it ensures that link bandwidth availability is not exceeded due to flow allocations.
    
    \item \label{step2} the second step checks whether the paths selected at \ref{step1} also guarantee the \textit{maximum} latency budget (i.e., $\cd{LatencyBudget} + \cd{Th}$)  when accounting for queuing delays caused by multiple flows traversing the same node.
    
\end{enumerate}

\noindent Steps \ref{step1} and \ref{step2} correspond one-to-one to the predicates \cd{possiblePaths/1} and \cd{validPaths/2} (lines 2--3) of the following Prolog code, respectively.
The variable \cd{Alloc} contains the result of \ref{step1} and the variable \cd{Alloc} the final result of \ref{step2}.

\begin{codenum}[firstnumber=1]
    dglbf(FinalAlloc) :-
        possiblePaths(Alloc), 
        validPaths(Alloc, FinalAlloc).
\end{codenum}

\noindent
For the sake of brevity and readability, we illustrate here the main checks performed by our declarative prototype in steps \ref{step1} and \ref{step2}, and refer our readers to \cite{wscc} for further details.

At \ref{step1}, for each \cd{flow(F, SrcN, DstN)} to be placed, the plain version of \toolname checks the following predicate with respect to the allocation \cd{Alloc} of flows currently being built:

\begin{codenum}
    eligiblePath(F, Alloc, (F, P, Path, NewMinB, Delay)) :-
       flow(F, SrcN, DstN),
       flowReqs(F, PktSize, _, Rate, LatencyBudget, Th),
       MinBudget is LatencyBudget - Th,
       candidate(P, SrcN, DstN, Path),
       pathOk(Path, MinBudget, Alloc, PktSize, Rate, NewMinB),
       additionalDelay(NewMinB, Path, Delay).
\end{codenum}

\noindent
Particularly, it retrieves the requirements associated with flow request \cd{F} (line 6), computes the minimum latency budget \cd{MinBudget} accounting for the set tolerance threshold \cd{Th} (line 7), and non-deterministically selects a candidate \cd{Path} from the pre-computed ones (line 8).

In turn, predicate \cd{pathOk/7} scans the selected path to check whether it features enough bandwidth and to update the residual minimum budget hop-by-hop (line 9). To do so, it repeatedly checks the following predicate at each hop \cd{N}--\cd{M} along the path:

\begin{codenum}[numbers=right]
    hopOk(N, M, DProp, BW, Alloc, PktSize, Rate, MinB, NewMinB) :- 
        node(M, DProc), 
        usedBandwidth(N, M, Alloc, UsedBW), 
        BW > UsedBW + Rate,
        DTransm is PktSize/BW,
        NewMinB is MinB - DProc - DTransm - DProp.
\end{codenum}

\noindent The \cd{hopOk/8} predicate computes the bandwidth \cd{UsedBW} consumed by previously placed flows as per \cd{Alloc} (line 13), and ensures the link between nodes \cd{N} and \cd{M} still has enough capacity to support the \cd{Rate} of the new flow (line 14). Last, it computes the residual latency budget \cd{NewMinB} by subtracting the node processing delay \cd{DProc}, the transmission delay \cd{DTransm} and the propagation delay \cd{DProp} (line 15-16) from the currently available budget \cd{MinB} for the considered hop.

Whenever a candidate path (if any) is found, the \cd{additionalDelay/3} predicate (line 10) checks whether the residual minimum delay is positive. If so, it computes the \cd{Delay} value to be enforced at each hop on the path to guarantee the minimum latency budget for each packet of the considered flow to reach its destination node (viz. \cd{LatencyBudget - Th}).
The result of step \ref{step1} is a list \cd{Alloc} made of tuples \cd{(F, P, Path, MinB, Delay)} that associate flow identifiers \cd{F} to path identifiers \cd{P}, along with the sequence of traversed hops \cd{Path}, the computed residual minimum budget \cd{NewMinB} and the additional \cd{Delay} to be configured. Such delay is computed by dividing the value \cd{NewMinB}\footnote{Additional delay is introduced whenever $\fncd{NewMinB} > 0$, i.e., when packets in a flow are expected to arrive before the minimum latency budget.} by the number of hops in the \cd{Path}.

At \ref{step2}, predicate \cd{validPaths/2} estimates the maximum end-to-end delay of each flow and checks that the paths selected at \ref{step1} for each flow do not introduce too long queuing delays that forbid packets to reach their destination node within their maximum latency budget.
This is performed by recursively checking that the following predicate holds for each flow allocation from \ref{step1}:

\begin{codenum}[numbers=right]
    compatiblePaths([AllocF|AFs], Paths, [FinalAllocF|FAFs]) :-
        AllocF = (F, P, Path, MinB, Delay),
        flowReqs(F, PktSize, BstSize, _, _, Th), 
        totQTime(Path, F, P, PktSize, BstSize, Paths, TotQTime),
        MaxB is MinB + 2*Th - TotQTime, MaxB >= 0,
        FinalAllocF is (F, P, Path, (MinB,MaxB), Delay),
        compatiblePaths(AFs, Paths, FAFs).
    compatiblePaths([], _, []).
\end{codenum}

Particularly, for each flow \cd{F}, \cd{compatiblePaths/3} retrieves its requirements in terms of \cd{PktSize} and \cd{BstSize} (line~19), and computes an estimate of the total maximum queuing time along the selected path \cd{P} (line 20). This accounts for the fact that a single node \cd{N} may be traversed by the currently considered flow \cd{F} \textit{and} by a set $\mathcal{F}_{\mathfncd{N}}$ of other flows. The following formula~\cite{glbf} estimates the queuing delay \cd{TotQTime} experienced by the last packet of a burst of flow \cd{F}, preceded by the bursts of the other flows in $\mathcal{F}_{\mathfncd{N}}$, which must be serialised on the output link with bandwidth \cd{BW}:

\begin{equation*}
  \frac{(\mathcd{F.BstSize}-1)*\mathcd{F.PktSize} + \sum_{\mathcd{G} \in \mathcal{F}_{\mathfncd{N}}} \mathcd{G.BstSize} * \mathcd{G.PktSize}}{\mathcd{BW}}\nonumber
\end{equation*}

The obtained value is then used to check whether the overall queuing time does not exceed the available maximum budget \cd{MaxB} for the considered flow (line 21). Said otherwise, it ensures that the result of subtracting the queuing time \cd{ToTQTime} from the \cd{MinB} computed at \ref{step1}, incremented by the toleration range \cd{2*Th}, is a non-negative value. If this check succeeds, the \cd{FinalAllocF} is added to the output list (lines 17 and 22).

\subsection{Extending \toolname with reliability}
\label{sec:rdglbf}

\noindent
We now illustrate how \toolname can be extended to deal not only with flows that have latency requirements to be guaranteed but also with flows that feature reliability requirements.
To this end, we rely on the following definitions:
\begin{itemize}
    \item \textit{path reliability} is an estimate of the probability that a path from source node $A$ to destination node $B$ is fully functional. To compute it, we rely on the reliability values $\cd{LinkRel}_\ell \in[0, 1]$ that annotate each link $\ell$ in the topology, and represent the probability that a link is functional (i.e. at any moment in time it does not incur a failure)~\cite{lin2015}. Given a path $\cd{P}$, the associated path reliability $r(\cd{P})$ is obtained by multiplying the reliability values of the links that make out the path, i.e. $r(\cd{P})=\prod_{\ell \in \cd{P}} \cd{LinkRel}_\ell$. Besides, if a set $R_{AB}$ of alternative protection paths is selected to route data from node \textit{A} to node \textit{B}, we can also compute route reliability as $\rho_{AB} = 1 - \prod_{\cd{P}\in R_{AB}} (1-r(\cd{P}))$ as in~\cite{lin2015}.
    \item \textit{path protection} foresees allocating two disjoint paths to a given flow (1+1 protected flow) so that, whenever one of the two paths incurs a failure, transmission might take place on the alternative one. Paths are considered as disjoint if they share only the source and destination nodes. 
    \item \textit{fate sharing avoidance} guarantees that two distinct flows (e.g. a data and a control flow) and their 1+1 replicas are routed through disjoint paths.
\end{itemize}

Equipping \toolname with reliability aspects requires extending the code of \Cref{sec:vanilla_dglbf} so as to include suitable declarative definitions of the above properties. In this section, we describe how those definitions can be implemented and used in our prototype by relying on the flexibility provided by its declarative nature. We hereinafter illustrate simple declarative definitions of the above properties that can be made more efficient or accurate in future releases of \toolname, as well as further extended to account for more constraints.

\subsubsection{Path reliability}
\label{sec:linkrel}

Including path reliability is obtained by adding predicate \cd{reliabilityOk/4} to the checks performed by predicate \cd{pathOk/7} during \ref{step1}. While recursively scanning, hop-by-hop, a selected candidate path, \cd{pathOk/7} can check the following predicate holds true:

\begin{codenum}[numbers=right]
    reliabilityOk(PathRel, LinkRel, ReqdRel, NewPathRel) :- 
        NewPathRel is PathRel * FeatRel, NewPathRel >= ReqRel.
\end{codenum}

\noindent
The \cd{reliabilityOk/4} predicate updates the value of path reliability while predicate \cd{pathOk/7} scans a selected path.  It computes the new value \cd{NewPathRel} for path reliability by multiplying the path reliability \cd{PathRel} computed up to the current link and the reliability \cd{LinkRel} of the link at hand. It incrementally checks the obtained result against the required reliability threshold \cd{ReqRel}, as specified in the \cd{reliabilityReqs/3} fact in the knowledge base (line 26).

\subsubsection{Path protection}
\label{sec:pathprotection}

Extending \toolname with path protection mechanisms requires to allocate two disjoint paths\footnote{Focussing on 1+1 path protection, we here show a simplified version of our code. Note that {\sf\scriptsize dgLBF} can actually search for an arbitrary set \cd{N} of disjoint paths onto which a flow can be replicated for reliability purposes.} for the flow to be protected. We thus extend \toolname so that whenever at least two candidate paths exists for the flow to be placed, the prototype attempts replicating the flow on two of them and checks that they are disjoint (except for their source and destination nodes). This is mainly achieved by changing the retrieval of a \cd{candidate/4} path (line 8) in \ref{step1} with the following \cd{validCandidate/4} predicate.

\begin{codenum}[numbers=right]
    validCandidate((S,D), P2, Path2, P1) :- 
        candidate(P2, S, D, Path2), dif(P, P1),
        pathProtection(Path2, P1).

    pathProtection(Path2, P1) :- 
        intermediateNodes(Path2, P), noIntersections(P, P1).
\end{codenum}

Indeed, \cd{validCandidate/4} retrieves a candidate path \cd{Path2} identified by \cd{P2} that interconnects a source node \cd{S} with a destination node \cd{D} and \texttt{dif}fers from the first path identified by \cd{P1} that has been chosen for routing the flow at hand (line 28). Subsequently, predicate \cd{pathProtection/2} retrieves intermediate nodes from the chosen candidate path \cd{Path2} and checks that they do not intersect with those of the previously selected candidate identified by \cd{P1} (line 31).

\subsubsection{Fate sharing avoidance}
\label{sec:nofatesharing}

Finally, to account for fate sharing avoidance among sets of flows, our prototype needs to be extended so as to enforce all specified \cd{antiAffinity/2} requirements. To do so, we further extend the predicate in charge of determining valid candidate paths onto which flows can be routed. Our new \cd{validCandidate/6} extends the one illustrated above (lines 27--29) as follows:

\begin{codenum}[numbers=right]
    validCandidate(F, (S,D), P, CPath, Ps, Out) :- 
        candidate(P, S, D, CPath), \textbackslash+ member(P, Ps),
        pathProtection(CPath, Ps), 
        noFateSharing(F, CPath, Out).

    noFateSharing(F, CPath, Out) :-
        allAntiAffinitiesPerFlow(F, AFPs),
        noFateSharing(CPath, AFPs).

    noFateSharing(CPath, [AFPath|AFPs]) :- 
        sharedFirstAndLast(CPath, AFPath, C, AF), 
        intersection(C, AF, []), 
        noFateSharing(CPath, AFPs).
    noFateSharing(_, []).
\end{codenum}

In addition to the previous check on path protection (line 34), this extended version relies on predicate \cd{noFateSharing/3} to ensure that, given a flow identified by \cd{F} and its candidate route \cd{CPath}, the flow placement does not violate anti-affinity constraints with other flows. Particularly, it retrieves all paths of already placed flows that feature anti-affinity constraint with \cd{F} (line 37). Through predicate \cd{noFateSharing/2} (line 39), it then recursively checks that the intersection between the current candidate path for flow \cd{F} and any of the paths chosen for flows in anti-affinity with \cd{F} is empty, viz. \cd{[]} (lines 40--41).

\subsubsection{Lessons learnt}
\label{sec:lessons}
To sum up, the declarative approach adopted in \toolname allowed us to flexibly extend our prototype so as to consider various important reliability aspects (viz. path reliability, path protection and fate sharing avoidance) when configuring network paths as per gLBF~\cite{glbf}. It was possible to do so by simply specifying high-level constraints and objectives as logical rules, without the burden of fully revising the modelling of the considered problem. It is worth mentioning that the above code constitutes a declarative, executable specification of a gLBF-based placement strategy enriched with reliability aspects, which can actually execute and output solution placements for sets of flows over a certain infrastructure. Besides, the prototype can be easily customised depending on specific needs, for instance by deciding to include only some of the checks on reliability (e.g. path protection, without fate sharing avoidance nor path reliability) or to account for alternative implementations of the proposed features.
\section{\toolname in action}
\label{sec:example}

\noindent
In this section, we first show \toolname at work on a simple example based on realistic data from the literature (\Cref{sec:usecase1}). Then, by relying on the same data, we perform a preliminary performance evaluation of our prototype considering average execution times and five different versions of \toolname (\Cref{sec:usecase2}), i.e. the plain version, one version accounting for all reliability properties, and three versions considering only one property each. 

\subsection{Use case: the Orion CEV topology}
\label{sec:usecase1}

\begin{figure}
    \centering
    \includegraphics[width=0.99\linewidth]{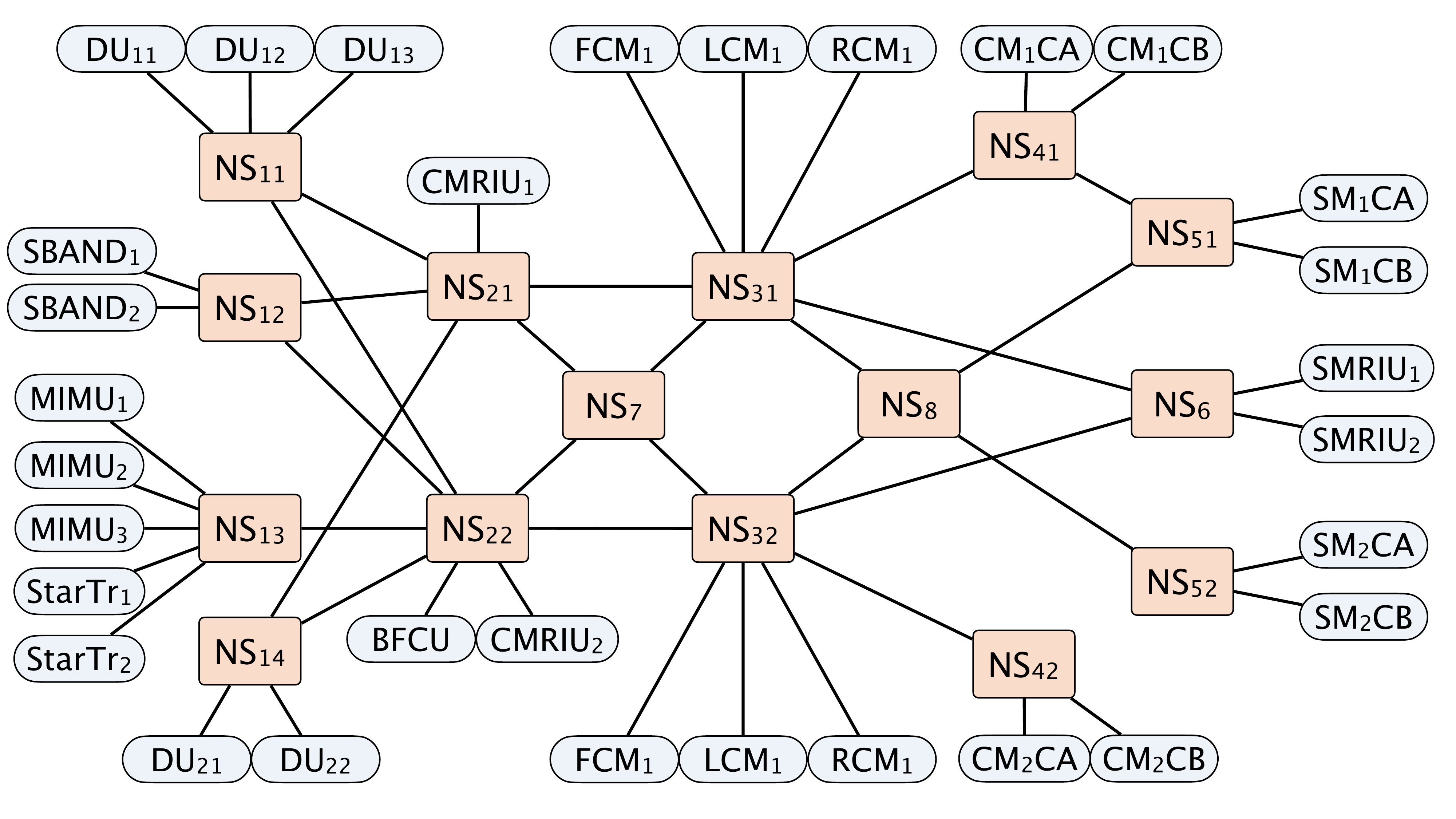}
    \caption{Orion CEV network topology~\cite{avb}.}
    \label{fig:orion-cev-topology}
\end{figure}

\noindent
This section demonstrates the application of \toolname to a simplified yet realistic scenario consisting of 3 flows to be placed onto an infrastructure based on the Orion Crew Exploration Vehicle (CEV) topology, depicted in~\cref{fig:orion-cev-topology}, and extensively used in prior research on TSN and industrial networking~\cite{tsnpeeper,avb, cev_icc, exp-setup}. Such a topology comprises 15 network switches (NS) and 31 end systems connected via 55 bidirectional links.

Some example of infrastructure nodes, flows, and reliability requirements are listed in \cref{code:example}. 
For the sake of conciseness we only show some example node and link facts\footnote{Full example data is available at: \url{https://github.com/di-unipi-socc/dgLBF/tree/main/sim/data}}, while flow requirements are fully listed.
For instance, node \cd{du11} features a \cd{3ms} latency budget and is connected to \cd{ns11} via a unidirectional link with \cd{3ms} latency, \cd{440Mpbs} bandwidth and a reliability value of \cd{99.28\%}.
The flow identified by \cd{f2}, starting from node \cd{ns12} and ending in \cd{ns6}, features a packet size of \cd{0.01Mb}, a burst size of \cd{4} packets, a bitrate of \cd{10Mbps} and a latency budget of \cd{100ms} with a \cd{10ms} toleration threshold, viz. it can tolerate delays between $90$ and $110$ ms. The other two streams \cd{f1} and \cd{f3} cannot share their path, thus they are subject to an anti-affinity constraint.

\begin{figure}[!ht]
    \begin{code}[gobble=4]
    node(du11, 3).      link(du11, ns11, 3, 440, 0.9928).
    node(du12, 4).      link(du12, ns11, 4, 421, 0.9913).
    \dots
    node(ns52, 4).      link(ns6, ns31, 2, 250, 0.9942).
    node(ns6, 4).       link(ns6, ns32, 3, 389, 0.9926).

    flow(f1, du11, sm2cb).  dataReqs(f1, 0.008, 3, 5, 50, 10).
    flow(f2, ns12, ns6).    dataReqs(f2, 0.01, 4, 10, 100, 10).
    flow(f3, du21, sm1cb).  dataReqs(f3, 0.008, 5, 6, 40, 10).

    reliabilityReqs(f1, 0.9, 1).
    reliabilityReqs(f2, 0.85, 2).
    reliabilityReqs(f3, 0.95, 1).

    antiAffinity(f1, [f3]).     antiAffinity(f3, [f1]).
    \end{code}
\caption{Example knowledge base (partial).}\label{code:example}
\end{figure}
Querying \toolname over the input in \cref{code:example} 
retrieves a solution to place each input flow, as shown in~\Cref{tab:output}. For instance, the flow identified by \cd{f2} is allocated to two disjoint paths \cd{ns12}--\cd{ns21}--\cd{ns31}--\cd{ns6} and \cd{ns12}--\cd{ns22}--\cd{ns32}--\cd{ns6}. For the first replica, as packets could take 75 ms less than the minimum required latency (15 ms $<$ 90 ms), \toolname suggests adding a delay of 25 ms per hop to be enforced via gLBF mechanisms (so that they arrive at exactly 90 ms). 
The retrieved \cd{MaxB} of 95 ms represents the maximum delay that a packet can tolerate and includes both the tolerance threshold (20 ms) and the remaining budget (75 ms), reduced by the queuing time that is negligible in our use case. Note that \cd{f1} and \cd{f3} come with entirely disjoint paths due to their anti-affinity constraint, and all the featured reliability values are higher than the required ones.


\begin{table*}[!ht]
\centering
\caption{Example output.}\label{tab:output}
\begin{tabular}{|c|c|c|c|c|c|c|}
\hline
\textbf{Flow} & \textbf{Replicas} & \textbf{Path} & \textbf{Path Rel. {[}\%{]}} & \textbf{MinB {[}ms{]}} & \textbf{MaxB {[}ms{]}} & \textbf{Per-hop additional delay {[}ms{]}} \\ \hline
f1 & 1 & {[}{\footnotesize \texttt{du11, ns11, ns22, ns32, ns8, ns52, sm2cb}}{]} & 97.04 & 7 & 27 & 1.17 \\ \hline
f2 & 1 & {[}{\footnotesize \texttt{ns12, ns21, ns31, ns6}}{]} & 98.32 & 75 & 95 & 25 \\ \hline
f2 & 2 & {[}{\footnotesize \texttt{ns12, ns22, ns32, ns6}}{]} & 98.15 & 71 & 91 & 23.7 \\ \hline
f3 & 1 & {[}{\footnotesize \texttt{du21, ns14, ns21, ns31, ns41, ns51, sm1cb}}{]} & 96.31 & 0 & 20 & 0 \\ \hline
\end{tabular}
\end{table*}

\subsection{Performance evaluation on the Orion CEV topology}
\label{sec:usecase2}
\noindent Always relying on the Orion CEV topology, we conducted a first set of experiments to evaluate the scalability of \toolname over a lifelike use case.
The assessment includes several versions of \toolname, each incorporating different features to handle specific constraints:
\begin{description}[font=\normalfont\itshape\xspace]
    \item[(Plain)] The base version of the tool~\cite{wscc}, which handles flow placement without considering reliability, path protection, or anti-affinity constraints. This version serves as a baseline for comparison.
    \item[(Path Protection)] This version introduces 1+1 path protection, ensuring that flows requiring it are assigned two disjoint paths. We evaluated it with two pairs of flows to test its effectiveness in providing redundancy.
    \item[(Anti-Affinity)] In this configuration, only anti-affinity constraints are considered, ensuring that a single pair of flows is routed independently to avoid fate-sharing.
    \item[(Reliability)] This version considers only cumulative reliability requirements computed by multiplying link reliability values along the selected path.
    \item[(All)] The most advanced version of \toolname, combining all the previous features into a single configuration.
\end{description}

For each configuration, we conducted the experiments using input batches of flows of increasing size: 150, 225, 300, 375, and 400 flows. Moreover, we evaluated each configuration over ten independent runs to account for variability in performance. Each run's input parameters, including flow specifications and topology details, are consistent with those described in~\cite{exp-setup}.

\begin{figure*}[!ht]
    \centering
    
    \begin{subfigure}{0.49\textwidth}
        \centering
        \includegraphics[width=\textwidth]{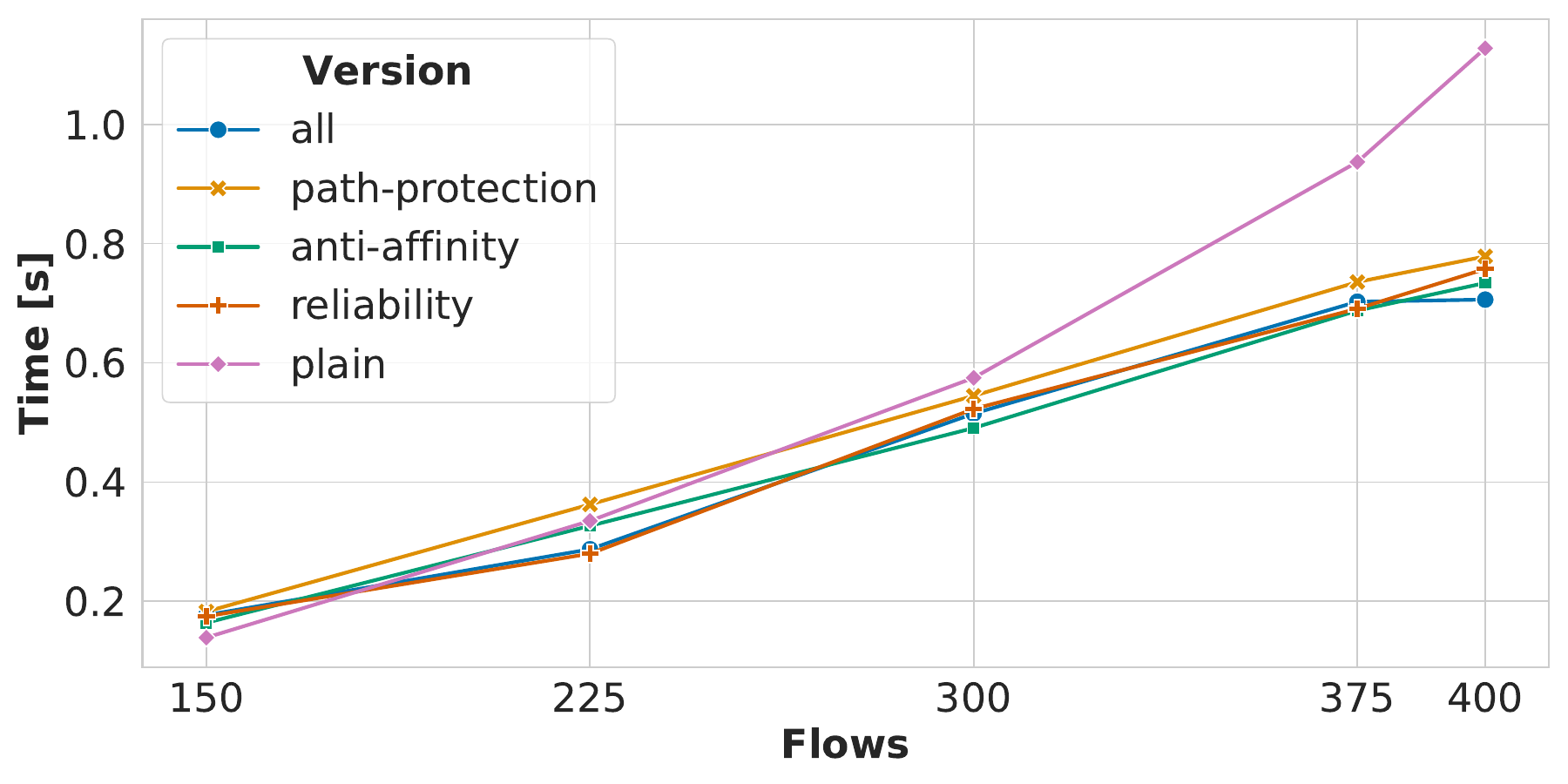}
        \caption{Execution times.}\label{fig:execution-times-cev}
    \end{subfigure}
    \hfill
    \begin{subfigure}{0.49\textwidth}
        \centering
        \includegraphics[width=\linewidth]{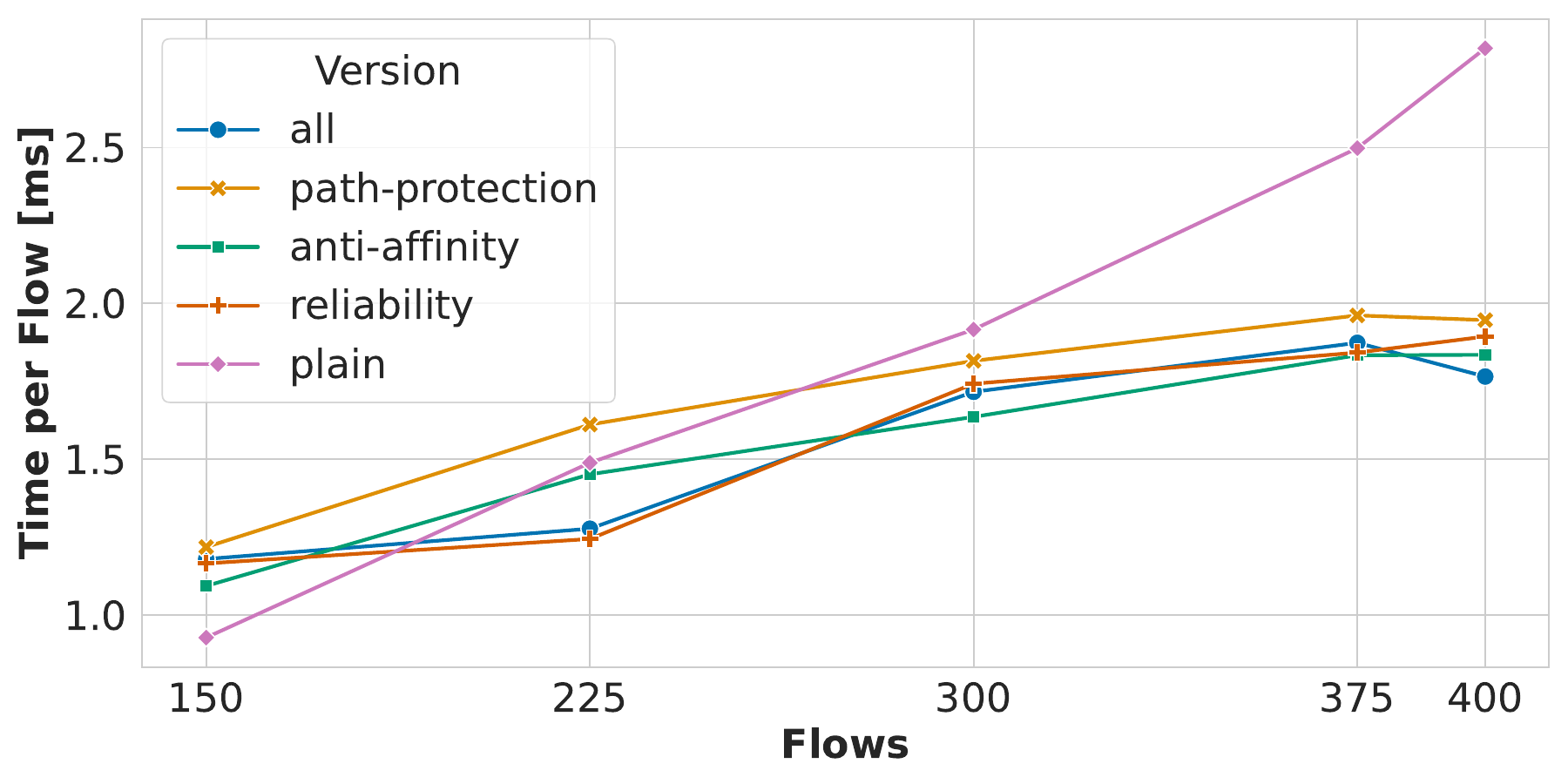}
        \caption{Time per flow.}\label{fig:time-per-flow-cev}
    \end{subfigure}

    \caption{Scalability assessment on CEV topology.}\label{fig:assessment-cev}
\end{figure*}

We present the obtained execution times in \cref{fig:assessment-cev}. \Cref{fig:execution-times-cev} illustrates the average execution times for each version of \toolname as the batch size of flows increases, while \cref{fig:time-per-flow-cev} the average time per flow placement across the same configurations and input sizes.

\Cref{fig:execution-times-cev} demonstrates that the execution times for all versions of \toolname exhibit quasi-linear growth concerning the input size, except for the \textit{Plain} version that shows a slightly exponential trend. For all versions, the execution time settles below 1.1 seconds, even for the largest batch size of 400 flows. 
As aforementioned, the \textit{Plain} version shows a slightly different growth in execution time due to the non-deterministic behaviour of the Prolog reasoner's implementation. Since this plain version does not benefit from additional constraints to reduce the search space (e.g., anti-affinity), it must backtrack more frequently during flow placement due to the Prolog's recursive decision-making process.

\Cref{fig:time-per-flow-cev} further supports these observations by analysing the average time required to place each flow. Like the total execution times, the \textit{Plain} version exhibits a higher increase in time per flow as the batch size grows. In contrast, the other versions maintain a quasi-linear behaviour. Notably, even for the largest batch size, the maximum time per flow remains extremely low across all configurations, reaching only 3 ms in the worst-case scenario. The time required to place each flow is up to 2 ms, for the most constrained version when handling 400 flows. This shows the efficiency of \toolname in managing individual flows, even handling all reliability constraints.

While the preliminary results show \toolname’s efficiency and scalability, they face limitations due to the limited path diversity of the Orion CEV topology, which often hinders satisfying path protection and anti-affinity requirements. Further and larger-scale experiments are reported in the next section.
\section{Scalability Assessment}
\label{sec:scalability}

\noindent
In this section, we further evaluate the performance of our prototype by conducting experiments
using two distinct models to generate random infrastructure graphs: the Barabasi-Albert (BA) and the Erdos-Renyi (ER) models. Besides, we only focus on the \textit{All} version of \toolname. 

\subsection{Experiment Setup}
\noindent
Experiments were run on a machine featuring Ubuntu 22.04.5 LTS (GNU/Linux 5.15.0-1065-nvidia x86\_64) equipped with 1.5TB of RAM and a 3.7GHz AMD EPYC 9534 64-Core processor, using SWI-Prolog 9.2.7~\cite{swipl} and Python 3.10.12\footnote{The code to run all experiments is publicly available at: \url{https://github.com/di-unipi-socc/dgLBF/tree/main/sim}.}.

\paragraph{Infrastructures}
We generated networks of $2^i$ nodes with $i \in [4,10]$, resulting in networks ranging from 16 to 1024 nodes. We used the BA model to create graphs with a preferential attachment mechanism, setting each node degree to $i$. In contrast, we configured the ER model with an edge attachment probability of 0.7, leading to denser networks. 
The connectivity patterns differentiate the two models, with ER graphs exhibiting random edge distribution and BA graphs displaying scale-free properties, where higher-degree nodes form hubs that can create a more clustered, hub-and-spoke-like structure.
All the scenarios consider a \cd{ProcessingTime} in the range $[1, 5]$ ms, with each link featuring a latency in $[1, 5]$ ms and a bandwidth in $[500, 1500]$ Mbps. These three parameters are randomly generated within their respective ranges, following a uniform distribution.

\paragraph{Flows} 
We randomly generated flows within each network topology to simulate different and realistic traffic scenarios. We varied the number of flows across a wide range, starting from 500 and increasing in steps of 500, up to a maximum of 10,000 flows per experiment. Among them, we selected $log_2(\#flows)$ random anti-affinity constraints as pairs of flows that must avoid fate sharing. 
We also used three probability values (25\%, 50\%, and 75\%) to determine whether a flow requires path protection.
Each flow originates from and is destined to randomly selected nodes within the network infrastructure. The \cd{PktSize} is set to $0.008$ Mb, and the \cd{BstSize} to $3$ packets. The \cd{Rate} is randomly generated within $[2, 8]$ Mbps, the \cd{LatencyBudget} within $[30, 60]$ ms, and the toleration threshold \cd{Th} within $[10, 20]$ ms, all following a uniform distribution.

\paragraph{Runs and time constraints} 
For each combination of topology type, infrastructure size and input requests (number of flows, probability value), we conducted ten independent runs -- with fixed and reproducible seeds -- to account for randomness in the graph generation and flow requirements' assignment.
Moreover, we limited each experiment with an 1800-second timeout (30 minutes) to limit the simulation duration and save computational resources.

\subsection{Experimental results}

\begin{figure*}[!ht]
    \centering
    
    \begin{subfigure}{0.32\textwidth}
        \centering
        \includegraphics[width=\textwidth]{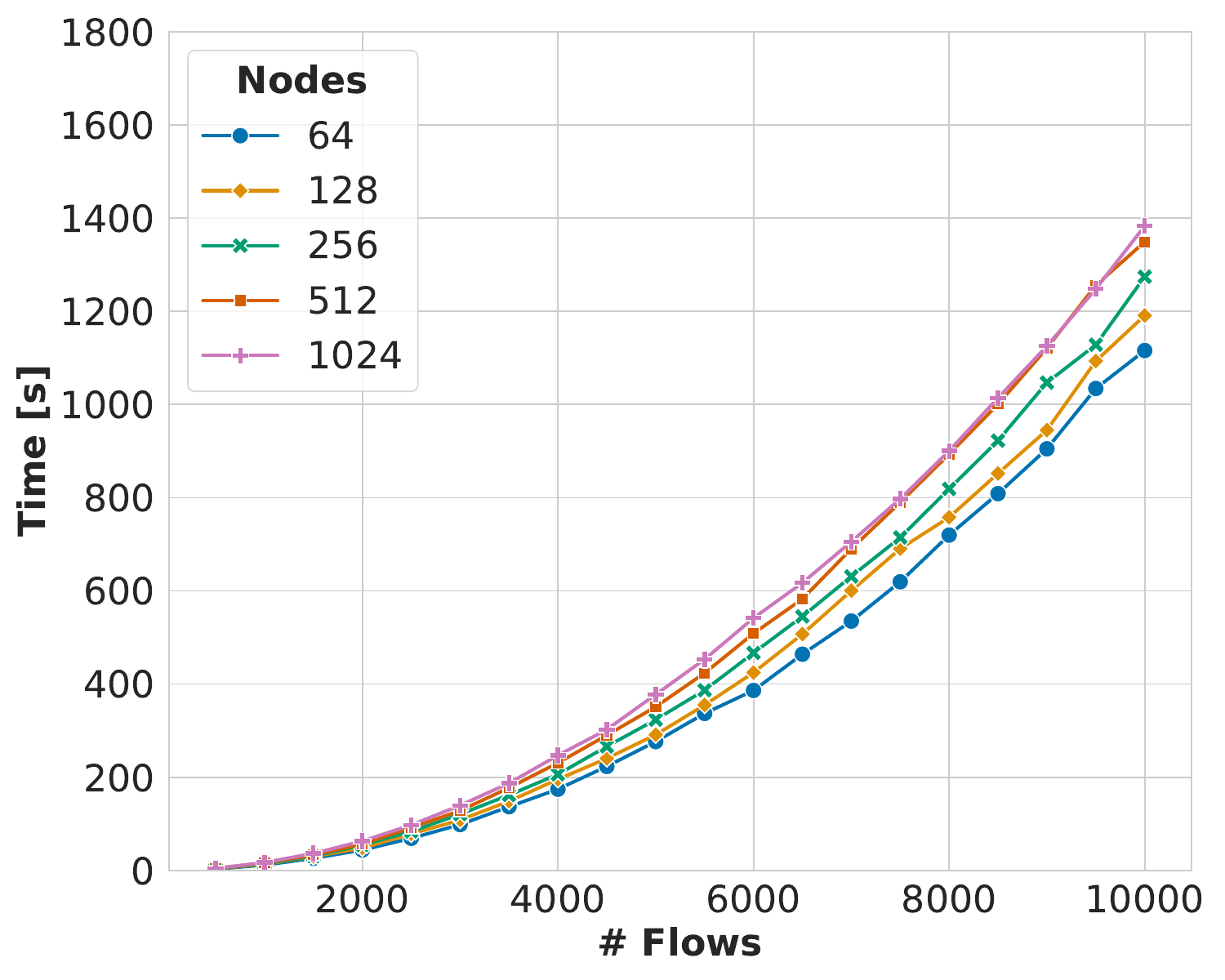}
        \caption{BA, 25\%}\label{fig:et_ba_25}
    \end{subfigure}
    \hfill
    \begin{subfigure}{0.32\textwidth}
        \centering
        \includegraphics[width=\textwidth]{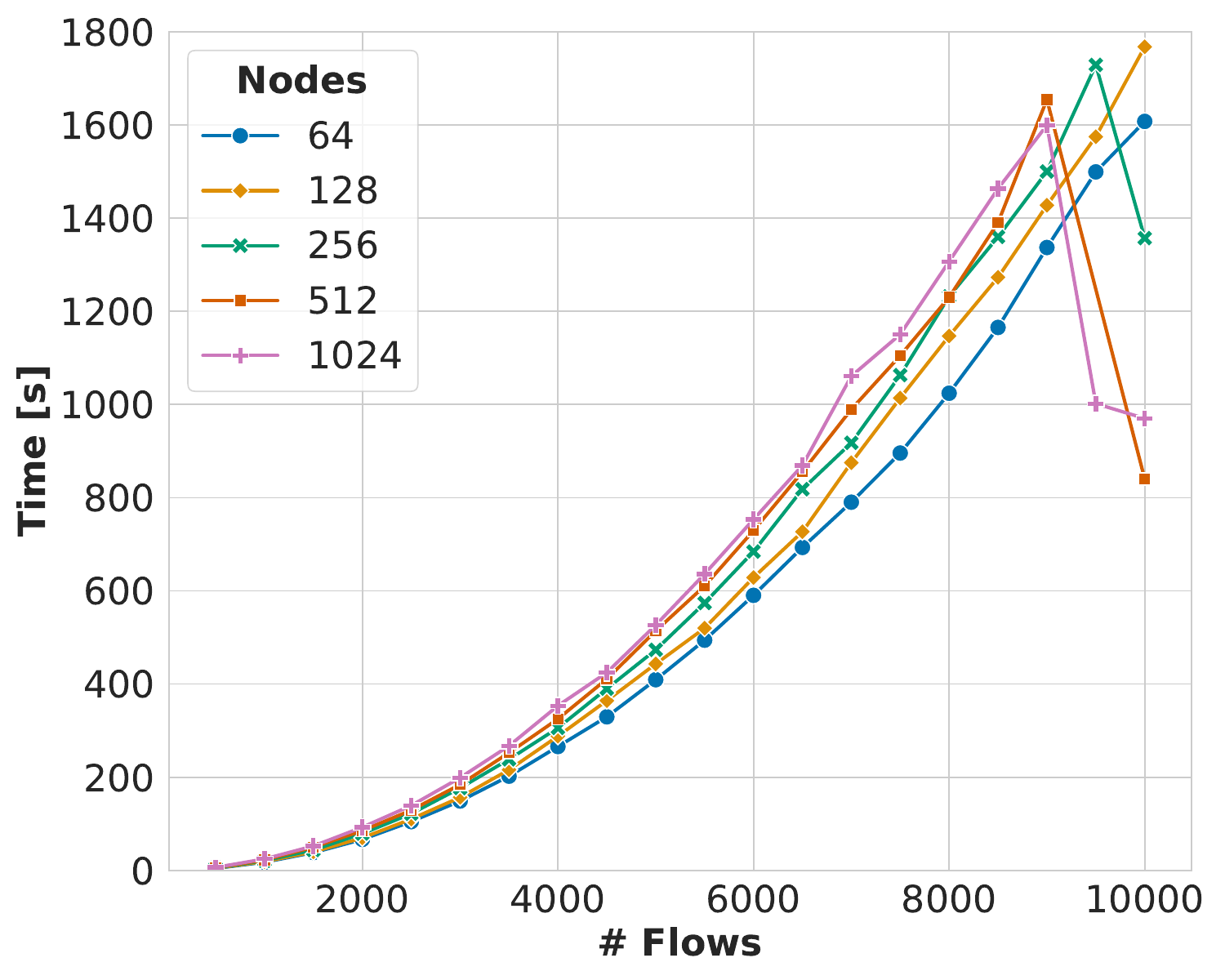}
        \caption{BA, 50\%}\label{fig:et_ba_50}
    \end{subfigure}
    \hfill
    \begin{subfigure}{0.32\textwidth}
        \centering
        \includegraphics[width=\textwidth]{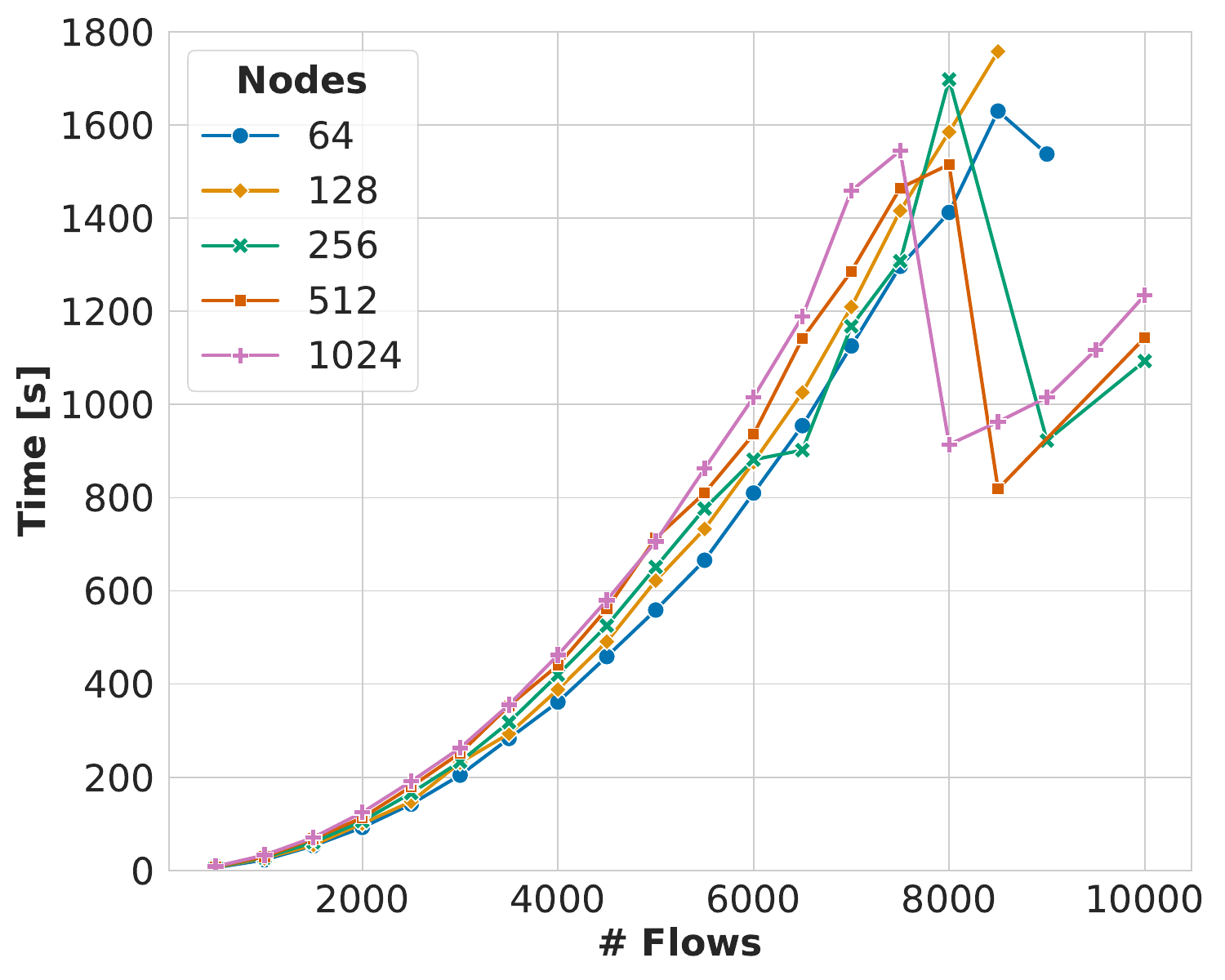}
        \caption{BA, 75\%}\label{fig:et_ba_75}
    \end{subfigure}
    \par\medskip
    \begin{subfigure}{0.32\textwidth}
        \centering
        \includegraphics[width=\textwidth]{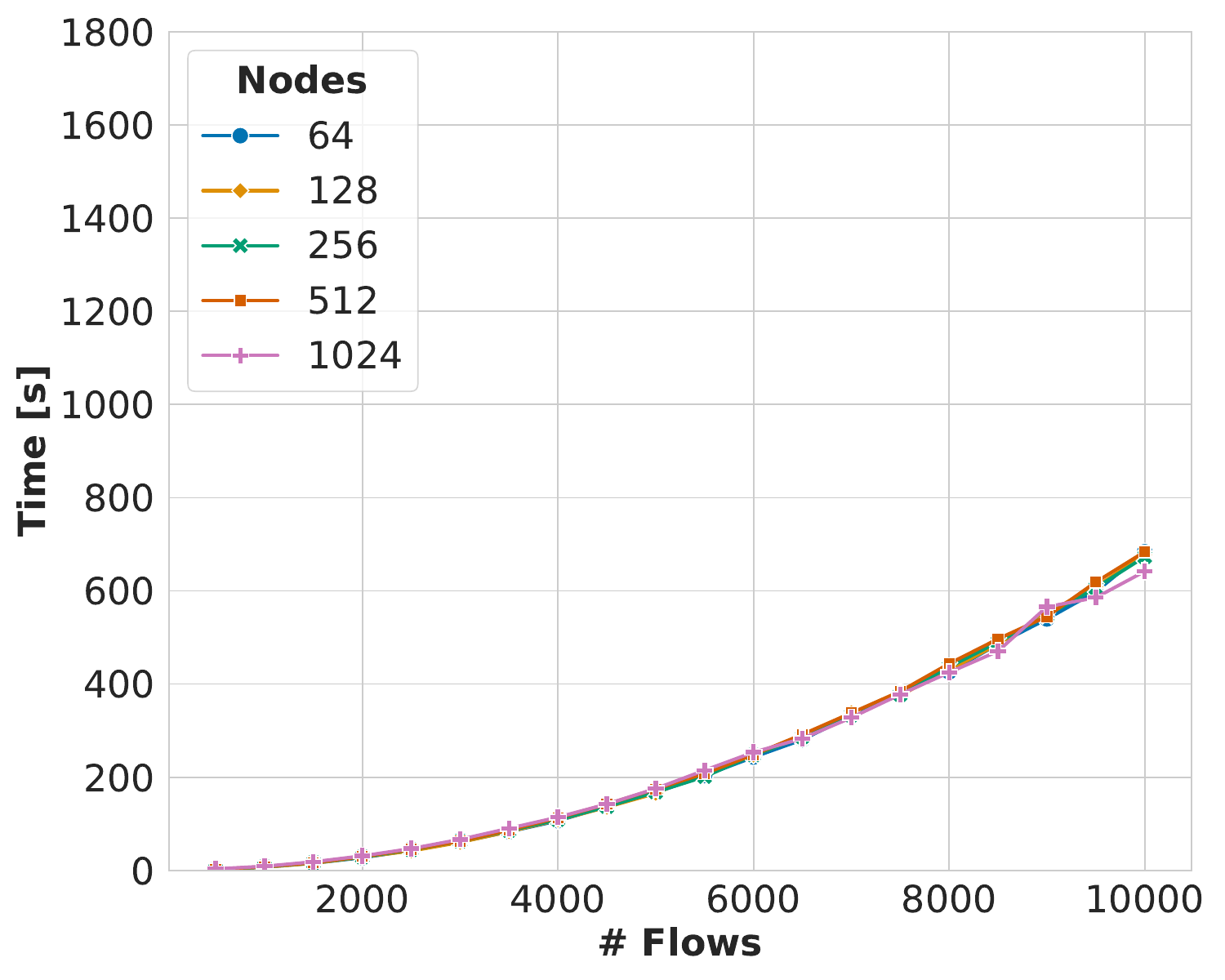}
        \caption{ER, 25\%}\label{fig:et_er_25}
    \end{subfigure}
    \hfill
    \begin{subfigure}{0.32\textwidth}
        \centering
        \includegraphics[width=\textwidth]{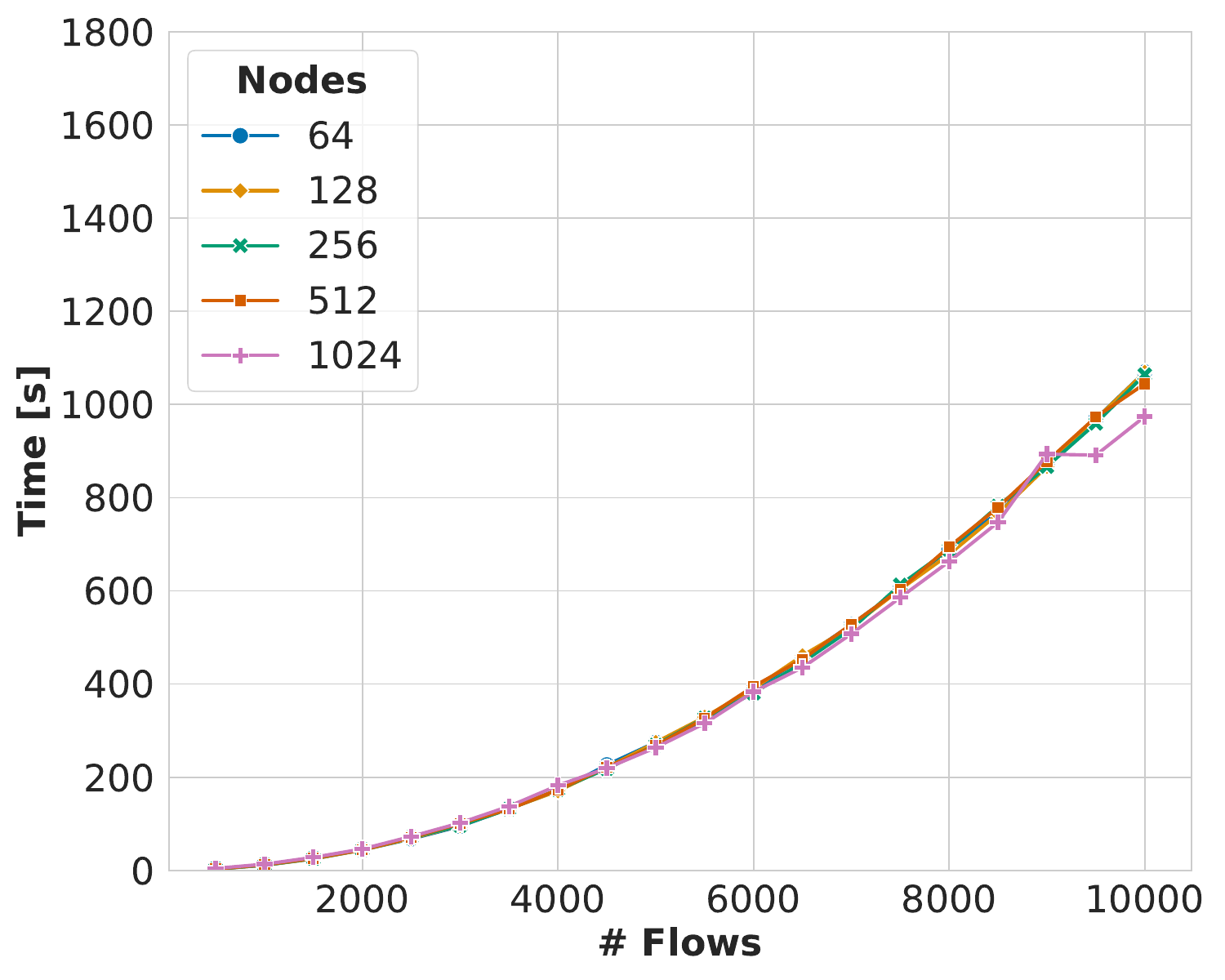}
        \caption{ER, 50\%}\label{fig:et_er_50}
    \end{subfigure}
    \hfill
    \begin{subfigure}{0.32\textwidth}
        \centering
        \includegraphics[width=\textwidth]{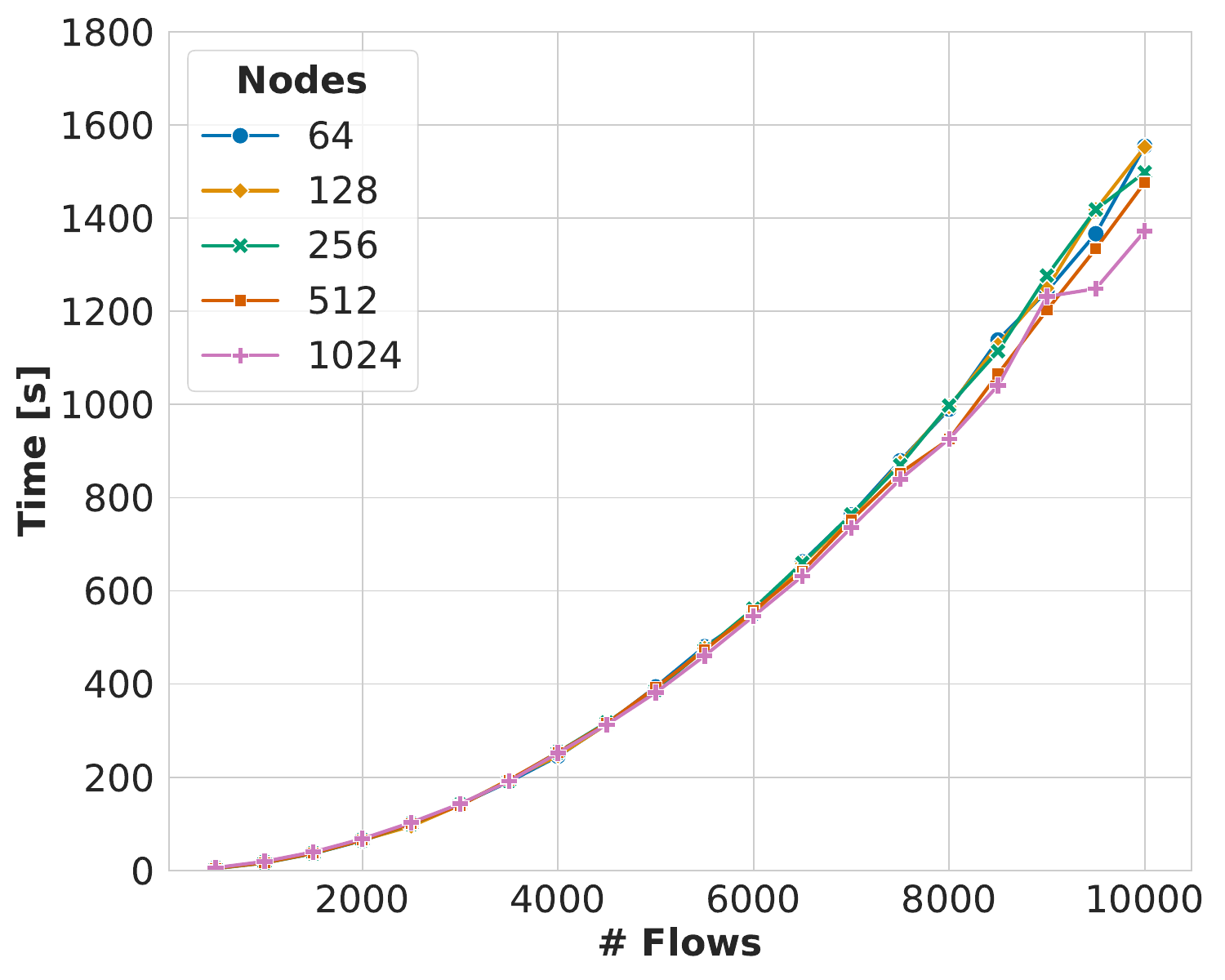}
        \caption{ER, 75\%}
        \label{fig:et_er_75}
    \end{subfigure}
    
    \caption{\textit{Execution times}' comparison for Barabasi-Albert (BA) and 
    Erdos-Renyi (ER) topologies under different path protection probabilities (25\%, 50\%, 75\%).}
    \label{fig:execution-times}
\end{figure*}

Hereafter, we present the experimental results to assess the computational efficiency and performance of \toolname under varying infrastructure sizes, flow loads, and replication probabilities. These tests aim to evaluate scalability limits, identifying the point at which the prototype can no longer find solutions within the allocated time.

\begin{figure*}[!ht]
    \centering
    \begin{subfigure}{0.32\textwidth}
        \centering
        \includegraphics[width=\textwidth]{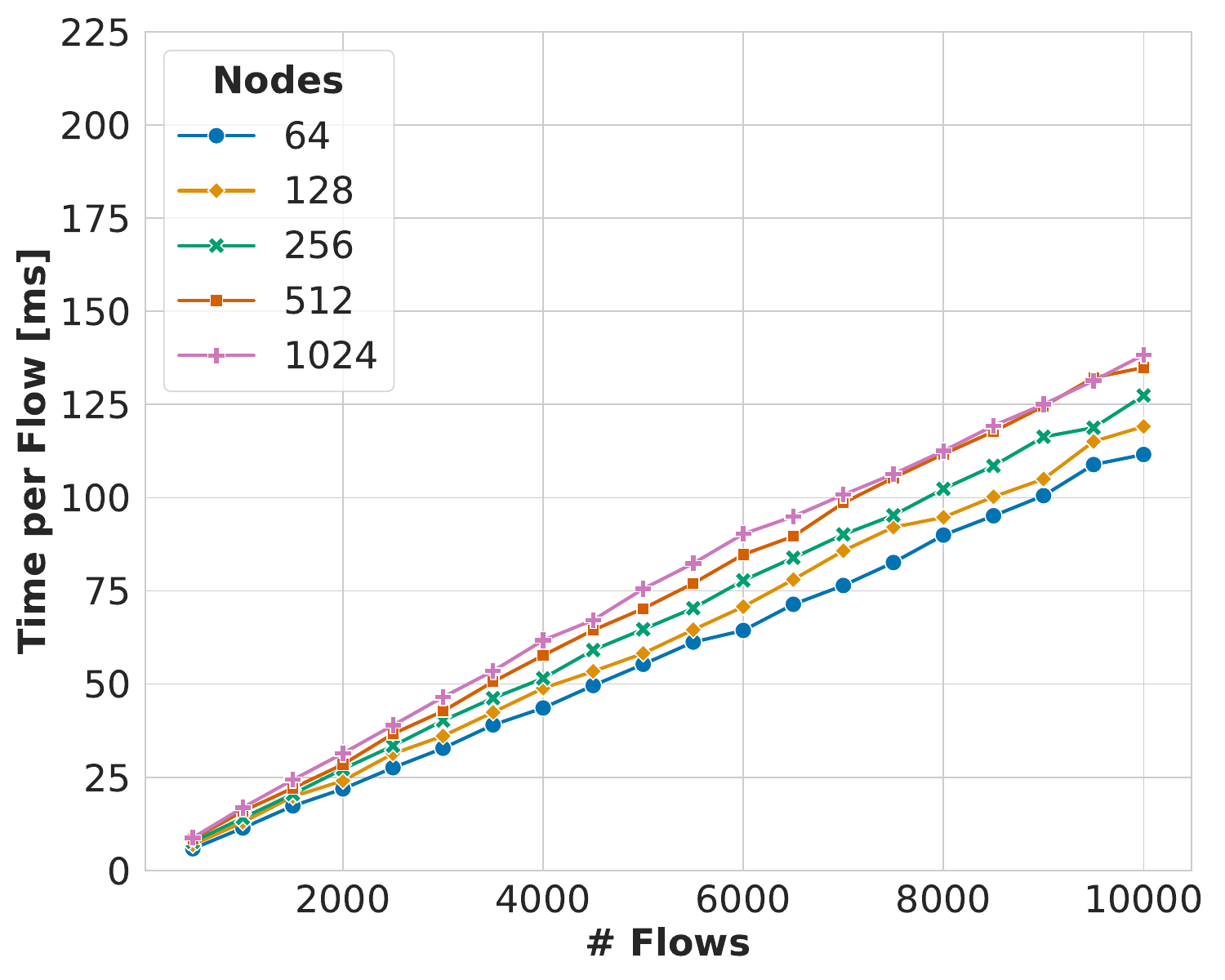}
        \caption{BA, 25\%}\label{fig:tpf_ba_25}
    \end{subfigure}
    \hfill
    \begin{subfigure}{0.32\textwidth}
        \centering
        \includegraphics[width=\textwidth]{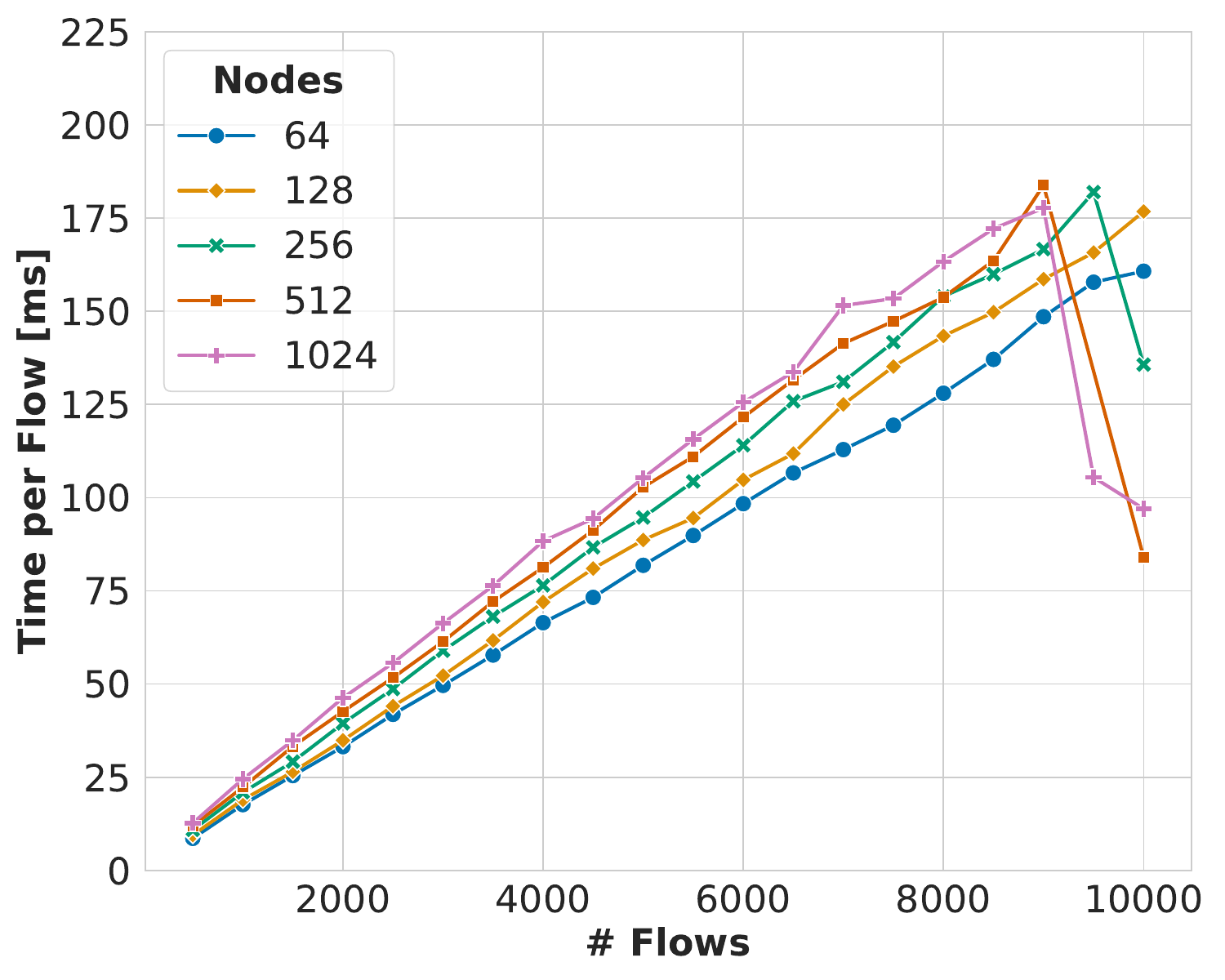}
        \caption{BA, 50\%}\label{fig:tpf_ba_50}
    \end{subfigure}
    \hfill
    \begin{subfigure}{0.32\textwidth}
        \centering
        \includegraphics[width=\textwidth]{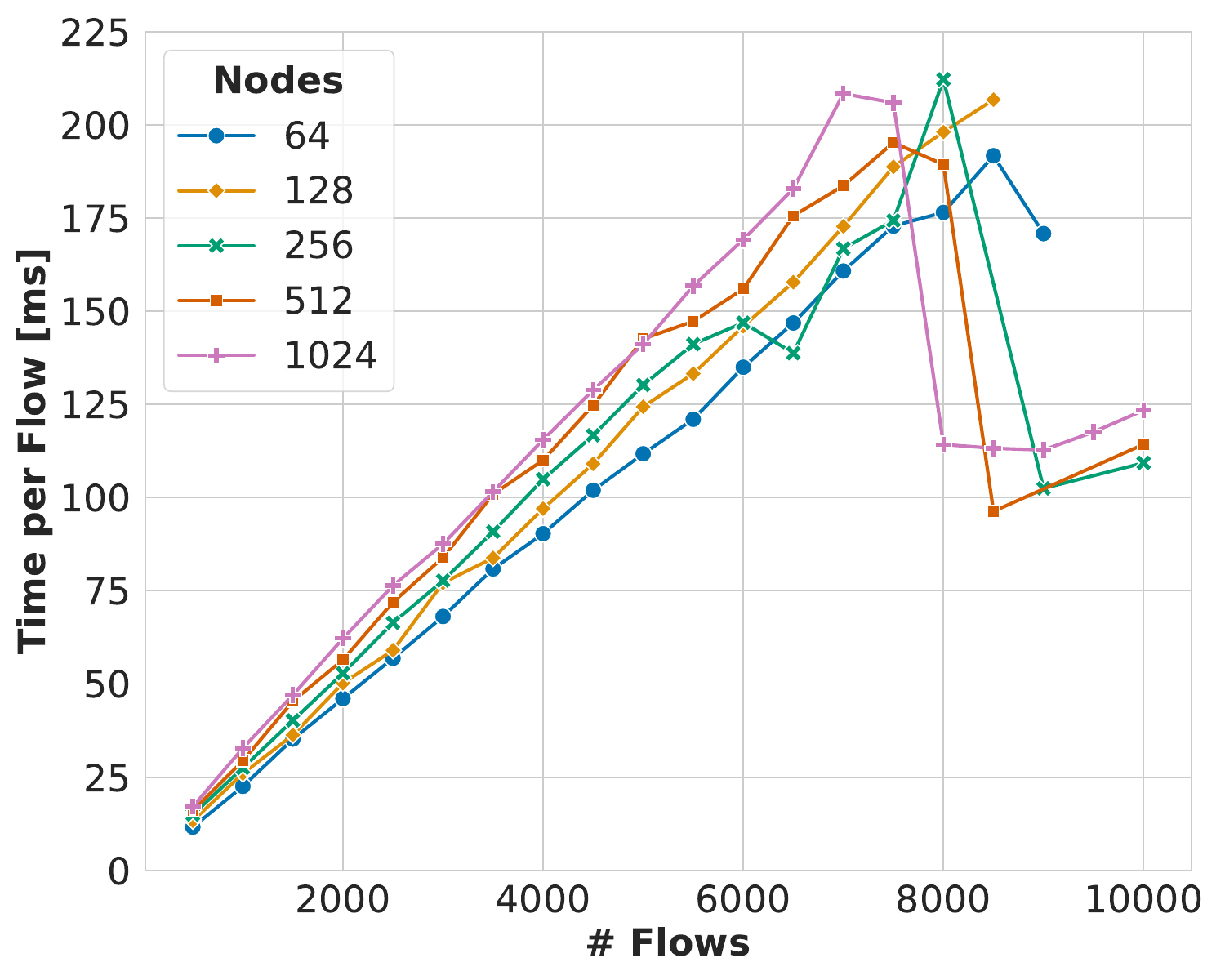}
        \caption{BA, 75\%}\label{fig:tpf_ba_75}
    \end{subfigure}
    \par\medskip
    \begin{subfigure}{0.32\textwidth}
        \centering
        \includegraphics[width=\textwidth]{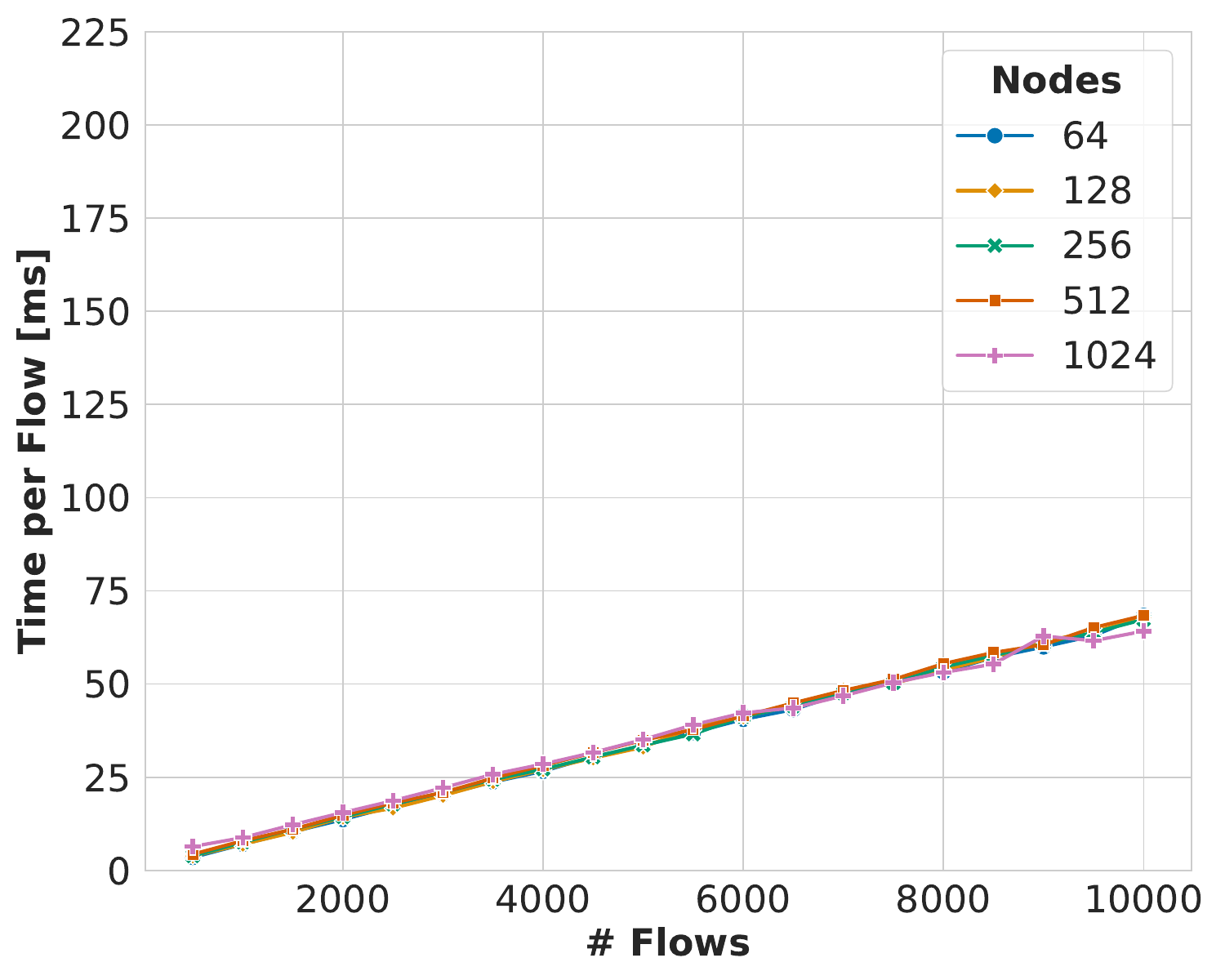}
        \caption{ER, 25\%}\label{fig:tpf_er_25}
    \end{subfigure}
    \hfill
    \begin{subfigure}{0.32\textwidth}
        \centering
        \includegraphics[width=\textwidth]{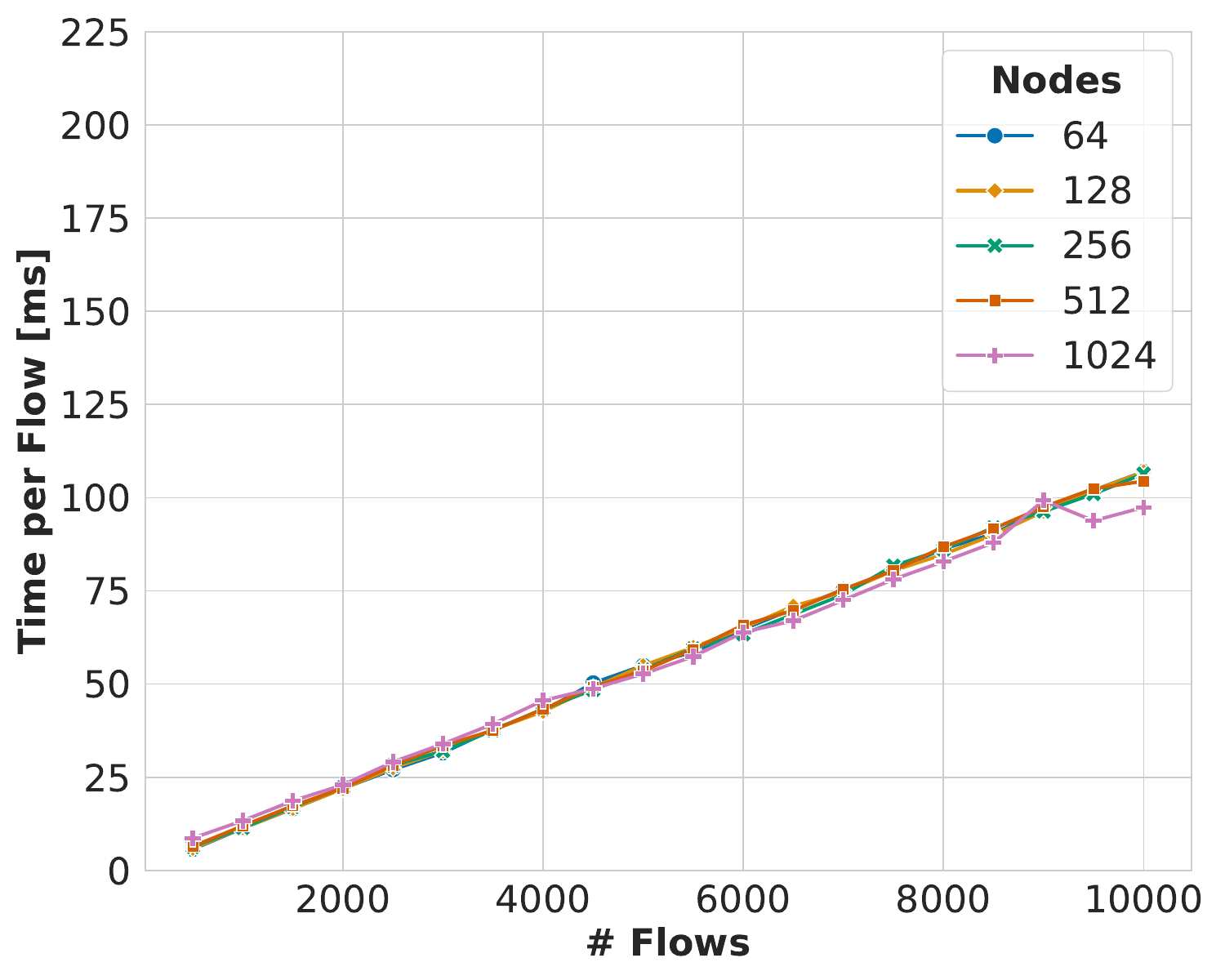}
        \caption{ER, 50\%}\label{fig:tpf_er_50}
    \end{subfigure}
    \hfill
    \begin{subfigure}{0.32\textwidth}
        \centering
        \includegraphics[width=\textwidth]{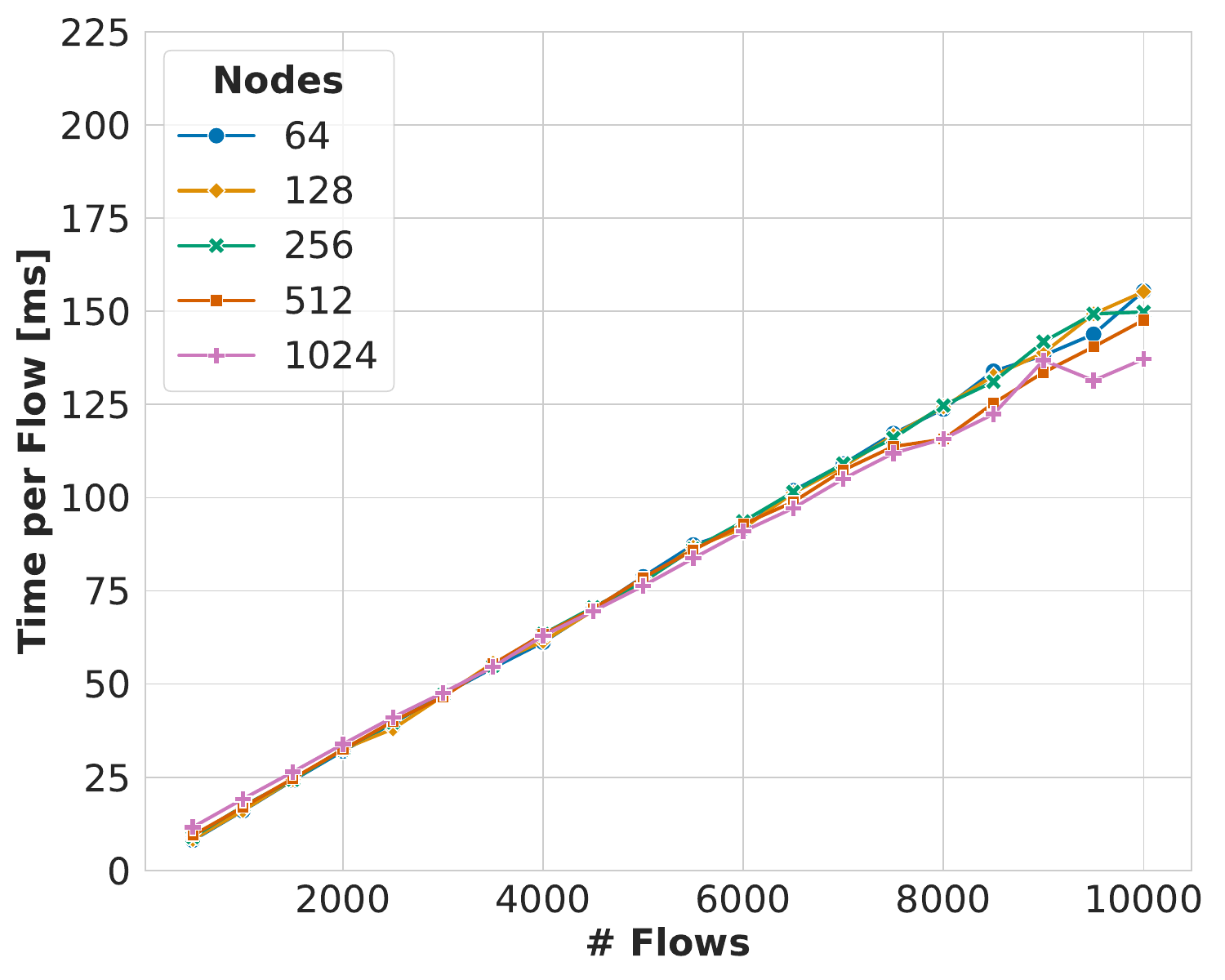}
        \caption{ER, 75\%}\label{fig:tpf_er_75}
    \end{subfigure}
    
    \caption{\textit{Time-per-flow}'s comparison for Barabasi-Albert (BA) and Erdos-Renyi (ER) topologies under different path protection probability (25\%, 50\%, 75\%).}
    \label{fig:time-per-flow}
\end{figure*}

We present the overall and per-flow execution times for the \textit{successful} runs in \cref{fig:execution-times,fig:time-per-flow}, respectively, highlighting the variation in computational efficiency when \toolname operates on the Barabasi-Albert (BA) and Erdos-Renyi (ER) topologies. As expected, increasing the number of nodes, flows, and path protection probabilities results in longer execution times. Larger infrastructure sizes and higher flow loads introduce greater complexity into the path computation process, leading to increased processing times.

\toolname exhibits higher execution times on BA than ER, which tends to approach the timeout limit of 1800 seconds under demanding scenarios. This behaviour is particularly evident at higher path protection probabilities (75\%), where the computational overhead is significantly higher on BA. Interestingly, execution times on BA also show slight variations depending on the infrastructure size, while in ER this is almost negligible except for extremely large flows and node inputs, where infrastructure size starts to impact results.

Beyond a certain scale -- approximately 8000 flows and especially on BA -- \toolname encounters timeouts when it fails to compute a valid solution within the time limit. This occurs because the Prolog reasoner backtracks extensively examining BA topologies, where fewer links lead to rapid saturation, making path computation increasingly complex. These timeouts cause execution to be terminated and are not included in average execution time calculations, leading to a visible drop in reported times in~\cref{fig:et_ba_50,fig:et_ba_75}, as only completed runs are accounted for. 
In contrast, \toolname remains stable on ER topologies, with timeouts only in the most complex cases.

The trend is even more evident in the time-per-flow analysis in \cref{fig:time-per-flow}. While execution time generally grows linearly with input size, higher flow loads lead to a drop in reported averages, always due to timeouts. This effect is more pronounced for BA, where fewer computations are completed within the time limit, while ER maintains a steadier increase, demonstrating greater resilience in handling large flow loads.

Despite the linear growth trends, time-per-flow remains within acceptable bounds across all configurations. The maximum reaches approximately 220 ms on BA and 155 ms on ER, even under the most demanding scenarios. This linearity in growth within each topology demonstrates \toolname’s capacity to efficiently manage individual flows, even as constraints and input sizes increase.

\subsection{Discussion}
\noindent The difference between the two considered topology types can be attributed to the ER model's more homogeneous and dense connectivity, where each node is more likely to be connected to other nodes. This increased connectivity reduces the complexity of finding alternative paths, thus lowering both the computational burden and the likelihood of failures, especially under high flow loads.
In contrast, the BA graphs, which rely on preferential attachment, tend to create a more hierarchical structure with hubs that dominate the network. While these hubs can efficiently handle small to medium traffic loads, they become bottlenecks under larger flow volumes, leading to longer execution times and higher failure rates. This effect is particularly evident when combined with high replication probabilities, as the need to find diverse paths around the hubs becomes computationally difficult, resulting in either timeouts or excessively long solution times.

Overall, it is worth mentioning that -- irrespectively of the considered topology type -- our prototype \toolname guarantees good scalability and manages to handle large-scale inputs, i.e. up to $\sim 6000$ flows on $1024$ nodes.

Lastly, considering the heaviest configuration that consistently produces results (Erdős-Rényi topology, 8000 flows, 75\% replication probability, 1024 nodes and $\simeq 95.000$ links), \toolname maintains low resource usage: CPU usage remains consistently around 0.75\% (ranging between 0.7\% and 0.8\%), while RAM consumption averages approximatively 800MB (700--840MB).

\section{Related Work}
\label{sec:related}


An overview of the current state of high-precision networking is provided by~\textcite{ClemmTNSM23}, which, in addition to outlining the main problems and existing solution approaches, identify as one of the key research challenges the need for precision-aware path steering and routing, enabling packets to be delivered following their respective service level guarantees.

To the best of our knowledge, this is the first study addressing bounded-delay routing with reliability constraints using gLBF, a promising candidate solution for implementing DetNet. In the following, we review related work in two key areas: routing with bounded delay guarantees and reliability within TSN and DetNet (\cref{sec:bounded-delay}); using declarative approaches through logic programming in networking (\cref{sec:decl-approacehs}).

\subsection{Routing with bounded delay guarantees and reliability}\label{sec:bounded-delay}

Most related work has focused on various routing and scheduling algorithms in layer-2 TSN. \textcite{min2023effective} consider the problem of multipath routing and Time Aware Shaper (TAS) scheduling in IEEE 802.1 TSNs and assume the use of frame replication and elimination for reliability (FRER) technique to guarantee reliable transmissions. Using FRER allows for relaxing the constraint of determining entirely disjoint paths. They propose a redundancy-weighted multipath routing algorithm and a heuristic scheduler based on member wait-and-forward constraints.
A similar problem is tackled by \textcite{routing_li_2024}, who relax timeslot scheduling constraints for replica flows to decrease computational complexity and improve the usage of resources in TSN-5G Networks.

Some work has also been done on routing for deterministic communication at layer 3 with the Deterministic Networking (DetNet) architecture.
\textcite{detnet_routing_drl24} address the problem of joint routing and scheduling for time-sensitive flows in a DetNet system. To this end, they propose a Deep Reinforcement Learning (DRL) framework to maximise the number of schedulable flows, optimise queue resource utilisation, and minimise end-to-end latency. Still, they do not consider traffic flow protection for reliability requirements. 
\textcite{detnet_protection_23} address the routing
and scheduling problem for protected traffic flows based on Cycle Specified Queuing and Forwarding
(CSQF). This involves selecting two disjoint paths between the source and destination
and generating reliable packet schedules along these paths. They model the problem using Integer Linear Programming (ILP) and propose two heuristic approaches (greedy and
Tabu-search). Since the authors assume the use of CSQF and Segment Routing, the proposed solution does not require per-flow state maintenance at each node but requires that all network nodes are synchronised within sub-microsecond accuracy. In contrast, thanks to the properties of gLBF, our work requires neither per-flow state maintenance at each node nor clock synchronisation while also allowing flexible specification of per-flow protection requirements ($1 + n$ disjoint paths, where $n$ may vary for each flow).

\subsection{Declarative Approaches in Networking}\label{sec:decl-approacehs}
A large body of work has been devoted to investigating the use of the declarative approach in networked systems.
Declarative networking~\cite{declarative_networking} is one of the first research efforts that leverages declarative, data-driven programming to specify and implement distributed protocols and services concisely~\cite{declarative_networking_mao}.
The original concept behind declarative networking is to use (recursive) database queries to specify nodes' routing tables declaratively, utilise a rule-based language derived from Datalog,  Network Datalog (NDlog), and perform distributed computation of shortest paths. Since then, declarative networking approaches have been applied in several networking applications, such as sensor networks~\cite{declarative_networking_sn}, dynamic overlay network compositions~\cite{declarative_networking_overlay}, fault tolerance~\cite{declarative_networking_ft}, configuration of network devices~\cite{declarative_networking_conf}, and verification of safety properties for SDN applications~\cite{declarative_networking_sdn}.

More recently, \textcite{switchlog} introduced SwitchLog, a logic programming language for real-world network switches. Inspired by NDlog-based approaches, it incorporates restrictions that enable the practical execution of line-rate switches. 

\textcite{declarative_networking_TN_23} propose NetSpec, a toolkit that synthesises formal specifications of network protocols in logic from input-output examples (e.g., the input may comprise facts about a network, while the output may comprise actual network state, such as the shortest path between two nodes). The specification language of NetSpec
 extends Datalog with negation, aggregation and user-defined functions.

Recently, we have applied logic programming to the management of the intent lifecycle in Intent-Based Network Systems,  focusing on two main areas:
\begin{enumerate*}[label=\textit{(\roman{*})}]
    \item the modelling and translation of intents for Virtual Network Function (VNF) service provisioning in a distributed edge-cloud infrastructure \cite{intent_chain_netsoft23}, and 
    \item the detection and resolution of conflicts within intent specifications \cite{intent_conflict_icin24}. 
\end{enumerate*}

All of the works above demonstrate, to varying extents, that expressing algorithms as logic programs offers several benefits. These include using a compact notation that can be easily extended to accommodate additional requirements and the ability to express programs in a form that is amenable to formal verification~\cite{declarative_networking_TN_23}.

However, to the best of our knowledge, only a few works have adopted declarative approaches, and more specifically, logic programming, for latency-bounded networking applications.~\textcite{prolog_tsn_2016} propose a network modelling approach based on logic programming for configuring and verifying in-vehicle TSN networks. The proposed model, which consists of Prolog facts and rules, aims to simplify the verification tasks of automotive network engineers by allowing them to focus solely on formulating the correct queries on the model. These queries are then solved using inference algorithms such as backward chaining.

\section{Concluding Remarks}
\label{sec:conclusions}

\noindent
In this article, we presented \toolname, a Prolog-based prototype for declarative traffic engineering, building upon the principles of guaranteed Latency-Based Forwarding (gLBF)~\cite{glbf} and extending them with reliability aspects. By integrating path reliability, path protection, and anti-affinity policies, \toolname addresses challenges in large and dynamic networks where deterministic latency guarantees are essential.

In particular, dgLBF concisely integrates functionalities typically associated with a Path Computation Engine (PCE) and a high-precision deterministic flow Admission Controller (AC). Its plain implementation consists of $\simeq 50$ lines of code, while the extended version -- supporting path reliability, path protection, and fate-sharing avoidance mechanisms -- expands to $\simeq 70$ lines. This compact size makes it easily maintainable and customisable. Tests show that \toolname can admit 400 flows in a maximum of 1.1 seconds (800 ms with reliability, protection and fate-sharing requirements) in the CEV topology, while tests on larger-scale topologies show a linear increase in execution time per flow as the input size increases up to about 8000 flows.

Future work will explore incorporating additional dimensions into \toolname, such as energy efficiency, security constraints, and multi-operator network domains. We also intend to support negotiation capabilities so that it can propose multiple alternative solutions that partially fulfil the requirements in the case of resource shortages. Last, we plan to evaluate \toolname performance in live testbed environments, further refining its applicability to real-world scenarios.

\section*{Acknowledgement}
This work was partially supported by the European Union - Next Generation EU under the Italian National Recovery and Resilience Plan (NRRP), Mission 4, Component 2, Investment 1.3, CUP C59J24000110004, partnership on “Telecommunications of the Future” (PE00000001 - program “RESTART”).
\AtNextBibliography{\footnotesize}
\printbibliography[heading=bibintoc]
\end{document}